\overfullrule=0pt

\newcount\mgnf

\mgnf=0
\ifnum\mgnf=0
      \def\openone{\leavevmode\hbox{\ninerm 1\kern-3.3pt\tenrm1}}%
      \def\*{\vglue0.3truecm} 
      \hsize=17truecm
      \vsize=23.truecm
      \parindent=4.pt
      \baselineskip=0.45cm
      \font\titolo=cmbx12 
      \font\titolone=cmbx10 scaled\magstep 2
      \font\cs=cmcsc10 
       
      \font\ottorm=cmr8

      \font\msytw=msbm10

      \font\indbf=cmbx10 scaled\magstep1 
      \fi

\ifnum\mgnf=1
      \def\openone{\leavevmode\hbox{\ninerm 1\kern-3.63pt\tenrm1}}%
      \def\*{\vglue0.5truecm}
      \magnification=\magstep1
      \hoffset=0.truecm
      \hsize=16truecm
      \vsize=24.truecm
      \baselineskip=1.4em  plus 0.05em minus0.1em 
      \parindent=1em
      \lineskip=0.1em\lineskiplimit=0.1em      
      \parskip=0.1pt plus1pt
      \font\titolo=cmbx12 scaled\magstep 1 
      \font\titolone=cmbx10 scaled\magstep 3 
      \font\cs=cmcsc10 scaled\magstep 1
      \font\ottorm=cmr8 scaled\magstep 1 
      
      \font\msytw=msbm10 scaled\magstep1 

      \font\indbf=cmbx10 scaled\magstep2 \fi

\global\newcount\numsec
\global\newcount\numapp
\global\newcount\numfor
\global\newcount\numfig
\global\newcount\numsub
\global\newcount\numlemma
\global\newcount\numtheorem
\global\newcount\numdef
\global\newcount\appflag 

\numsec=0\numapp=0\numfig=1

\def\veroparagrafo{\number\numsec}
\def\veraformula{\number\numfor}
\def\veraappendice{\number\numapp}
\def\verasub{\number\numsub}
\def\verafigura{\number\numfig}
\def\verolemma{\number\numlemma}
\def\verotheorem{\number\numtheorem}
\def\veradef{\number\numdef}

\def\section(#1,#2)
    {\advance\numsec by 1\numfor=1\numsub=1%
    \numlemma=1\numtheorem=1\numdef=1\appflag=0%
    \SIA p,#1,{\veroparagrafo} %
    \write15{\string\Fp (#1){\secc(#1)}}%
    \write16{ sec. #1 ==> \secc(#1)}%
    \hbox to \hsize{\titolo\hfill\number\numsec. #2\hfill%
    \expandafter{\alato(sec. #1)}}\*}

\def\appendix(#1,#2)
    {\advance\numapp by 1\numfor=1\numsub=1%
    \numlemma=1\numtheorem=1\numdef=1\appflag=1%
    \SIA p,#1,{A\veraappendice} %
    \write15{\string\Fp (#1){\secc(#1)}}%
    \write16{ app. #1 ==> \secc(#1)  }%
    \hbox to \hsize{\titolo\hfill Appendix A\number\numapp. #2\hfill%
    \expandafter{\alato(app. #1)}}\*}

\global\newcount\numsec\global\newcount\numapp
\global\newcount\numfor\global\newcount\numfig\global\newcount\numsub
\global\newcount\numlemma\global\newcount\numtheorem\global\newcount\numdef
\global\newcount\appflag \numsec=0\numapp=0\numfig=1
\def\veroparagrafo{\number\numsec}\def\veraformula{\number\numfor}
\def\veraappendice{\number\numapp}\def\verasub{\number\numsub}
\def\verafigura{\number\numfig}
\def\verolemma{\number\numlemma}
\def\verotheorem{\number\numtheorem}
\def\veradef{\number\numdef}

\def\section(#1,#2){\advance\numsec by 1\numfor=1\numsub=1%
\numlemma=1\numtheorem=1\numdef=1\appflag=0%
\SIA p,#1,{\veroparagrafo} %
\write15{\string\Fp (#1){\secc(#1)}}%
\write16{ sec. #1 ==> \secc(#1)  }%
\hbox to \hsize{\titolo\hfill \number\numsec. #2\hfill%
\expandafter{\alato(sec. #1)}}\*}

\def\appendix(#1,#2){\advance\numapp by 1\numfor=1\numsub=1%
\numlemma=1\numtheorem=1\numdef=1\appflag=1%
\SIA p,#1,{A\veraappendice} %
\write15{\string\Fp (#1){\secc(#1)}}%
\write16{ app. #1 ==> \secc(#1)  }%
\hbox to \hsize{\titolo\hfill Appendix A\number\numapp. #2\hfill%
\expandafter{\alato(app. #1)}}\*}

\def\senondefinito#1{\expandafter\ifx\csname#1\endcsname\relax}

\def\SIA #1,#2,#3 {\senondefinito{#1#2}%
\expandafter\xdef\csname #1#2\endcsname{#3}\else \write16{???? ma
#1#2 e' gia' stato definito !!!!} \fi}

\def \Fe(#1)#2{\SIA fe,#1,#2 }
\def \Fp(#1)#2{\SIA fp,#1,#2 }
\def \Fg(#1)#2{\SIA fg,#1,#2 }
\def \Fl(#1)#2{\SIA fl,#1,#2 }
\def \Ft(#1)#2{\SIA ft,#1,#2 }
\def \Fd(#1)#2{\SIA fd,#1,#2 }

\def\etichetta(#1){%
\ifnum\appflag=0(\veroparagrafo.\veraformula)%
\SIA e,#1,(\veroparagrafo.\veraformula) \fi%
\ifnum\appflag=1(A\veraappendice.\veraformula)%
\SIA e,#1,(A\veraappendice.\veraformula) \fi%
\global\advance\numfor by 1%
\write15{\string\Fe (#1){\equ(#1)}}%
\write16{ EQ #1 ==> \equ(#1)  }}

\def\getichetta(#1){Fig. \verafigura%
\SIA g,#1,{\verafigura} %
\global\advance\numfig by 1%
\write15{\string\Fg (#1){\graf(#1)}}%
\write16{ Fig. #1 ==> \graf(#1) }}

\def\etichettap(#1){%
\ifnum\appflag=0{\veroparagrafo.\verasub}%
\SIA p,#1,{\veroparagrafo.\verasub} \fi%
\ifnum\appflag=1{A\veraappendice.\verasub}%
\SIA p,#1,{A\veraappendice.\verasub} \fi%
\global\advance\numsub by 1%
\write15{\string\Fp (#1){\secc(#1)}}%
\write16{ par #1 ==> \secc(#1)  }}

\def\etichettal(#1){%
\ifnum\appflag=0{\veroparagrafo.\verolemma}%
\SIA l,#1,{\veroparagrafo.\verolemma} \fi%
\ifnum\appflag=1{A\veraappendice.\verolemma}%
\SIA l,#1,{A\veraappendice.\verolemma} \fi%
\global\advance\numlemma by 1%
\write15{\string\Fl (#1){\lm(#1)}}%
\write16{ lemma #1 ==> \lm(#1)  }}

\def\etichettat(#1){%
\ifnum\appflag=0{\veroparagrafo.\verotheorem}%
\SIA t,#1,{\veroparagrafo.\verotheorem} \fi%
\ifnum\appflag=1{A\veraappendice.\verotheorem}%
\SIA t,#1,{A\veraappendice.\verotheorem} \fi%
\global\advance\numtheorem by 1%
\write15{\string\Ft (#1){\thm(#1)}}%
\write16{ th. #1 ==> \thm(#1)  }}

\def\etichettad(#1){%
\inum\appflag=0{\veroparagrafo.\veradef}%
\SIA d,#1,{\veroparagrafo.\veradef} \fi%
\inum\appflag=1{A\veraappendice.\veradef}%
\SIA d,#1,{A\veraappendice.\veradef} \fi%
\global\advance\numdef by 1%
\write15{\string\Fd (#1){\defz(#1)}}%
\write16{ def. #1 ==> \defz(#1)  }}

\def\Eq(#1){\eqno{\etichetta(#1)\alato(#1)}}
\def\eq(#1){\etichetta(#1)\alato(#1)}
\def\eqg(#1){\getichetta(#1)\alato(fig #1)}
\def\sub(#1){\0\palato(p. #1){\bf \etichettap(#1)\hskip.3truecm}}
\def\lemma(#1){\0\palato(lm #1){\cs Lemma \etichettal(#1)\hskip.3truecm}}
\def\theorem(#1){\0\palato(th #1){\cs Theorem \etichettat(#1)%
\hskip.3truecm}}
\def\definition(#1){\0\palato(df #1){\cs Definition \etichettad(#1)%
\hskip.3truecm}}

\def\equv(#1){\senondefinito{fe#1}$\clubsuit$#1%
\write16{eq. #1 non e' (ancora) definita}%
\else\csname fe#1\endcsname\fi}
\def\grafv(#1){\senondefinito{fg#1}$\clubsuit$#1%
\write16{fig. #1 non e' (ancora) definito}%
\else\csname fg#1\endcsname\fi}

\def\secv(#1){\senondefinito{fp#1}$\clubsuit$#1%
\write16{par. #1 non e' (ancora) definito}%
\else\csname fp#1\endcsname\fi}

\def\lmv(#1){\senondefinito{fl#1}$\clubsuit$#1%
\write16{lemma #1 non e' (ancora) definito}%
\else\csname fl#1\endcsname\fi}

\def\thmv(#1){\senondefinito{ft#1}$\clubsuit$#1%
\write16{th. #1 non e' (ancora) definito}%
\else\csname ft#1\endcsname\fi}

\def\defzv(#1){\senondefinito{fd#1}$\clubsuit$#1%
\write16{def. #1 non e' (ancora) definito}%
\else\csname fd#1\endcsname\fi}

\def\equ(#1){\senondefinito{e#1}\equv(#1)\else\csname e#1\endcsname\fi}
\def\graf(#1){\senondefinito{g#1}\grafv(#1)\else\csname g#1\endcsname\fi}
\def\secc(#1){\senondefinito{p#1}\secv(#1)\else\csname p#1\endcsname\fi}
\def\lm(#1){\senondefinito{l#1}\lmv(#1)\else\csname l#1\endcsname\fi}
\def\thm(#1){\senondefinito{t#1}\thmv(#1)\else\csname t#1\endcsname\fi}
\def\defz(#1){\senondefinito{d#1}\defzv(#1)\else\csname d#1\endcsname\fi}
\def\sec(#1){{\S\secc(#1)}}

\def\BOZZA{
\def\alato(##1){\rlap{\kern-\hsize\kern-1.2truecm{$\scriptstyle##1$}}}
\def\palato(##1){\rlap{\kern-1.2truecm{$\scriptstyle##1$}}}
}

\def\alato(#1){}
\def\galato(#1){}
\def\palato(#1){}

{\count255=\time\divide\count255 by 60
\xdef\hourmin{\number\count255}
        \multiply\count255 by-60\advance\count255 by\time
   \xdef\hourmin{\hourmin:\ifnum\count255<10 0\fi\the\count255}}

\def\oramin{\hourmin }

\def\data{\number\day/\ifcase\month\or gennaio \or febbraio \or marzo \or
aprile \or maggio \or giugno \or luglio \or agosto \or settembre
\or ottobre \or novembre \or dicembre \fi/\number\year;\ \oramin}
\setbox200\hbox{$\scriptscriptstyle \data $}
\footline={\rlap{\hbox{\copy200}}\tenrm\hss \number\pageno\hss}

\let\a=\alpha \let\b=\beta  \let\g=\gamma     \let\d=\delta  \let\e=\varepsilon
  \let\h=\eta    
   \let\l=\lambda
\let\m=\mu    \let\n=\nu            \let\p=\pi      \let\r=\rho
\let\s=\sigma \let\t=\tau   \let\f=\varphi     \let\c=\chi
\let\ps=\psi   \let\o=\omega 
 \let\D=\Delta     \let\L=\Lambda  
           
\let\O=\Omega 

\def\\{\hfill\break} \let\==\equiv

\let\io=\infty 

\let\0=\noindent

\def\der{\hbox{\rm d}}
 
\let\bs=\backslash

\def\tende#1{\,\vtop{\ialign{##\crcr\rightarrowfill\crcr
 \noalign{\kern-1pt\nointerlineskip}
 \hskip3.pt${\scriptstyle #1}$\hskip3.pt\crcr}}\,}
\def\otto{\,{\kern-1.truept\leftarrow\kern-5.truept\to\kern-1.truept}\,}

\def\hp{{\hat\psi}}
\def\PP{{\cal P}}\def\EE{{\cal E}}\def\VV{{\cal V}}
\def\HH{{\cal H}}\def\WW{{\cal W}}
\def\TT{{\cal T}}\def\NN{{\cal N}}\def\BB{{\cal B}}
\def\RR{{\cal R}}\def\LL{{\cal L}}
\def\DD{{\cal D}}

\def\der{{\rm d}}
\def\T#1{{#1_{\kern-3pt\lower7pt\hbox{$\widetilde{}$}}\kern3pt}}
\def\VVV#1{{\underline #1}_{\kern-3pt
\lower7pt\hbox{$\widetilde{}$}}\kern3pt\,}
\def\W#1{#1_{\kern-3pt\lower7.5pt\hbox{$\widetilde{}$}}\kern2pt\,}

\def\indica{\leaders \hbox to 0.5cm{\hss.\hss}\hfill}
\def\guida{\leaders\hbox to 1em{\hss.\hss}\hfill}
\mathchardef\oo= "0521

\def\pp{{\bf p}}\def\xx{{\bf x}}
\def\yy{{\bf y}}\def\kk{{\bf k}}
\def\zz{{\bf z}}\def\uu{{\bf u}}
 \def\bP{{\bf P}}\def\rr{{\bf r}}
\def\tt{{\bf t}}

\def\Halmos{\hfill\vrule height6pt width4pt depth2pt \par\hbox to \hsize{}}

\def\oo{{\underline \omega}}

\def\xxx{{\underline\xx}}

\def\qed{\raise1pt\hbox{\vrule height5pt width5pt depth0pt}}

\def\indic{\hbox{\raise-2pt \hbox{\indbf 1}}}

\def\RRR{\hbox{\msytw R}}

\def\lft{\left}
\def\rgt{\right}
\def\la{{\langle}}
\def\ra{{\rangle}}
\def\defi{{=}}

%
%
%
\def\ins#1#2#3{\vbox to0pt{\kern-#2 \hbox{\kern#1 #3}\vss}\nointerlineskip}
%
%
%
\newdimen\xshift \newdimen\xwidth \newdimen\yshift

\def\insertplot#1#2#3#4#5{\par%
\xwidth=#1 \xshift=\hsize \advance\xshift by-\xwidth \divide\xshift by 2%
\yshift=#2 \divide\yshift by 2%
\line{\hskip\xshift \vbox to #2{\vfil%
#3 \includegraphics{#4.pst}}\hfill \raise\yshift\hbox{#5} }}

%
%
%
\def\ins#1#2#3{\vbox to0pt{\kern-#2 \hbox{\kern#1 #3}\vss}\nointerlineskip}
%
%
%
\newdimen\xshift \newdimen\xwidth \newdimen\yshift

\def\insertplot#1#2#3#4#5{\par%
\xwidth=#1 \xshift=\hsize \advance\xshift by-\xwidth \divide\xshift by 2%
\yshift=#2 \divide\yshift by 2%
\line{\hskip\xshift \vbox to #2{\vfil%
#3 \includegraphics{#4.pst}}\hfill \raise\yshift\hbox{#5} }}

\openin14=\jobname.aux \ifeof14 \relax \else
\input \jobname.aux \closein14 \fi
\openout15=\jobname.aux

\font\tenmib=cmmib10
\font\sevenmib=cmmib10 scaled 800
\font\titolo=cmbx12
\font\titolone=cmbx10 scaled\magstep 2

\font\journal=cmti10
\font\pagine=cmti10

\font\cs=cmcsc10

\font\ninerm=cmr9
\font\ottorm=cmr8
\textfont5=\tenmib
\scriptfont5=\sevenmib
\scriptscriptfont5=\fivei

\font\msytw=msbm9 scaled\magstep1

\font\indbf=cmbx10 scaled\magstep2

\newskip\ttglue
\font\ottorm=cmr8\font\ottoi=cmmi8\font\ottosy=cmsy7
\font\ottobf=cmbx7\font\ottott=cmtt8\font\ottosl=cmsl8\font\ottoit=cmti7
\font\sixrm=cmr6\font\sixbf=cmbx7\font\sixi=cmmi7\font\sixsy=cmsy7
\font\fiverm=cmr5\font\fivesy=cmsy5\font\fivei=cmmi5\font\fivebf=cmbx5

\def\ottopunti{\def\rm{\fam0\ottorm}\textfont0=\ottorm%
\scriptfont0=\sixrm\scriptscriptfont0=\fiverm\textfont1=\ottoi%
\scriptfont1=\sixi\scriptscriptfont1=\fivei\textfont2=\ottosy%
\scriptfont2=\sixsy\scriptscriptfont2=\fivesy\textfont3=\tenex%
\scriptfont3=\tenex\scriptscriptfont3=\tenex\textfont\itfam=\ottoit%
\def\it{\fam\itfam\ottoit}\textfont\slfam=\ottosl%
\def\sl{\fam\slfam\ottosl}\textfont\ttfam=\ottott%
\def\tt{\fam\ttfam\ottott}\textfont\bffam=\ottobf%
\scriptfont\bffam=\sixbf\scriptscriptfont\bffam=\fivebf%
\def\bf{\fam\bffam\ottobf}\tt\ttglue=.5em plus.25em minus.15em%
\setbox\strutbox=\hbox{\vrule height7pt depth2pt width0pt}%
\normalbaselineskip=9pt\let\sc=\sixrm\normalbaselines\rm}

\openin14=\jobname.aux \ifeof14 \relax \else
\input \jobname.aux \closein14 \fi
\openout15=\jobname.aux

\def\chapter(#1,#2){{}}
\frenchspacing
\centerline{\titolone Nonperturbative Adler-Bardeen Theorem}
\vskip1cm
\centerline{{\titolo Vieri Mastropietro}} 
\vskip3mm \centerline{Dipartimento di Matematica, Universit\`a di Roma ``Tor Vergata''} 
\centerline{via della Ricerca Scientifica, I-00133, Roma} 
\leftskip=1cm
\rightskip=1cm
\vskip1cm
\0{\cs Abstract.} {\it The Adler-Bardeen theorem 
has been proved only as a statement valid at all orders in perturbation theory, without any
control on the convergence of the series.
In this paper  
we prove a nonperturbative version
of the Adler-Bardeen theorem in $d=2$ by using recently
developed technical tools in the theory of Grassmann integration.
}
\leftskip=0cm
\rightskip=0cm
\vskip.5cm
\0{\cs Keywords} Nonperturbative merthods; Chiral anomaly; resummation of the perturbation expansion;
Adler-Bardeen theorem; constructive QFT.
\vskip.5cm
\0{\cs PACS numbers} 11.15.Tk, 11.30.Rd,11.40.Ha,12.38Cy,11.10.Kk
\*
\*
\section(1, Introduction and Main results)

\vskip.5cm
\sub(1.1aklx){\it Anomalies in QFT.}
The chiral anomaly appears as a quantum correction
to the conservation of the axial current for massless fermions.
A crucial property is the 
{\it anomaly nonrenormalization}, which says that the chiral anomaly
is given {\it exactly} by its lower order contribution.
This property was proved first for $QED_4$
in [AB] in the well known  {\it Adler-Bardeen Theorem}: it was shown that
there is a dramatic cancellation, if a suitable regularization
is assumed, among the infinite collection
of Feynmann graphs contributing to the anomaly
and at the end it turns out that the anomaly
is given by a single graph (the famous "triangle graph"):
the result can be condensed
by the formula
$$\partial^\m j^5_\m={\a_0\over 4\pi}\e_{\m,\n,\r,\s}F^{\m,\n}
F^{\r,\s}\Eq(aa1)$$
where $\a_0$ is the unrenormalized coupling constant. 
Different proofs of \equ(aa1) were given later in [Z]
and [LS]: as the results in [AB], they were 
statements {\it 
valid at all orders in perturbation theory} and
with no control on the convergence of the series
itself. 
The property of the anomaly nonrenormalization
holds also in the Electroweak model where it plays a crucial role even to prove the renormalizability;
as the gauge fields couple to chiral currents,
the chiral anomaly would break the renormalizability, but a remarkable cancellation
between anomalies (not renormalized according the Adler-Bardeen theorem)
of different fermion species saves the theory and gives a confirmation
of the fermionic family structure as well. 

Recent textbooks tend to present
the anomaly nonrenormalization in a functional integral approach in which, following
the elegant treatment of [F], one recovers it from the Jacobian associated
to a chiral transformation. However,
as explained for instance in [A1], such methods {\it cannot be considered} simpler proofs
of the Adler-Bardeen theorem: the methods in [F]
essentially treat the gauge fields as {\it classical fields} 
so that they produce essentially {\it one loop results} and
eventual higher orders correction would be in any case not included.
Hence it is the validity of such functional approach which is justified by [AB]
rather than the contrary.

As the anomaly nonrenormalization is a quite delicate property,
against which several objections has been raised
along the years (see for instance [JJ], [AI] or [DMT]),
it would be desirable to go beyond perturbation theory. This seems
actually far from the present analytical possibilities in $d=4$, 
for the difficulty of giving
a real non-perturbative meaning to the functional integrals
expressing the theory; it is worth then to consider 
$d=2$ QFT models which have proven fruitful laboratories
to test general properties. 

In $d=2$ the perturbative analysis in [AB]
can be repeated with no essential modifications, see [GR], 
and still the anomaly nonrenormalization holds
in the form (in the Euclidean case) $\partial_\m j^5_\m=-i e\a\e^{\m,\n}\partial_\m A_\n$,
where $e$ is the coupling and $\a$ is the value of the 
"bubble graph" (replacing the "triangle  graph" in $d=4$).
It holds $\a={1\over 4\pi}$ or $\a={1\over 2\pi}$
depending if {\it dimensional} or {\it momentum} regularization is used.
Again nonperturbative informations cannot be obtained
by such a procedure, based on explicit cancellations between Feynmann
graphs. It is also claimed that
the anomaly nonrenormalization in $d=2$ can be derived
by an exact functional approach, see for instance [FSS];
indeed integrating out the fermions it turns out that
the partition function for many $d=2$ $QFT$ models
can be written as
$$
\int P(d A) {{\rm det}(\g_\m[\partial_\m+A_\m])\over
{\rm det}(\g_\m\partial_\m)}\Eq(12)
$$
where $A_{\m,\xx}=(A_{0,\xx},A_{1,\xx})$ are fields with
Gaussian measure $P(dA)$
with covariance $<A_{\m,\xx} A_{\n,\yy}>=e^2 \d_{\m,\n}v(\xx-\yy)$.
A similar expression holds for the generating functional.
It is well known [Se] that, {\it under suitably regularity conditions
over $A_\m$}, $\log {\rm det}(\g_\m\partial_\m+\g_\m A_\m)-\log {\rm det}(\g_\m \partial_\m)$
is quadratic in $A_\m$; replacing the determinant 
with a quadratic exponential one gets easily,by an explicit integration 
of the Gaussian integrals, that the anomalies are
{\it not renormalized by higher orders}. 
However in the above derivation {\it an approximation is implicit};
the fermionic determinant in \equ(12)
is given by a quadratic expression {\it only if} $A_\m$ is sufficiently regular,
but in \equ(12) {\it the integral is over all possible fields $A$},
hence one is
{\it neglecting} the contributions from the irregular fields.
A peculiarity of $d=2$ QFT is the existence
of some exact solutions; indeed it has been claimed [GR]
that the Adler-Bardeen theorem 
finds a nonperturbative verification 
from comparison with the operatorial
exact solution of [J],[K] in the case of 
contact current-curent interaction. However the regularization 
in the functional integrals or in 
the operatorial exact solution are different, hence
there is no guarantee [GL] that the Schwinger functions obtained
from functional integrals converge, removing cutoffs and in 
the massless limit, to the exact ones
(indeed this is not the case). In conclusion, even
in $d=2$ there are no rigorous verification of the Adler-Bardeen theorem
in a functional integral approach to QFT
beyond perturbation theory.

The rigorous construction of $d=2$ QFT models from functional integrals
is in general not trivial at all, as they 
appear to be related to the continuum limit of the correlations of 
coupled bidimensional Ising or vertex models 
[GM1], which are in general hard to compute [B].
Some $d=2$ QFT models has been 
deeply investigated in the Eighties in the framework 
of Constructive QFT (see [GK],[Le]),
and in recent times new powerful methods has been developed
in [BM], overcoming the well
known technical problem posed by the combination of 
a nonperturbative setting based on multiscale analysis [P],[G]
with the necessity of exploiting cancellations due to
local gauge symmetries. These new technical tools allow us to rigorously investigate,
{\it for the first time}, the properties of anomalies 
of $d=2$ QFT models constructed from functional integrals;
in particular, we can prove a {\it non-perturbative} version of the Adler-Bardeen 
under suitable conditions on the bosonic propagator, avoiding completely Feynmann
graphs expansions and with full rigor. 
\vskip.5cm
\sub(2.1aklx){\it Euclidean $QFT_2$.}
We consider an Euclidean QFT
in $d=1+1$ whose Schwinger function can be obtained from the following
functional integral
$$e^{\WW_{N,L}(J,\phi)}=
\int 
P_N(d\psi) P(d A)
e^{\int d\xx [e \bar\psi_\xx(A_{\m,\xx}\g^\m)\psi_\xx+
J_{\m,\xx}A_{\m,\xx}+
{\phi_\xx\bar\psi_\xx\over\sqrt{Z}}+
{\bar\phi_\xx\psi_\xx
\over \sqrt{Z}}]}
\Eq(1)$$
where $\phi,J$ are external fields,
$Z$ is the {\it wave function renormalization} and:

-)in $\L=[0,L]\times [0,L]$ a lattice $\L_{a}$ is introduced
whose sites are given
by the space-time points $\xx=(x,x_0)=(n a,n_0 a)$ 
with ${L/ a}$ integer and $n,n_0=-L/2 a,1,\ldots,L/2a-1$.
We also consider the set $\DD$ of space-time momenta 
$\kk=(k,k_0)$
with $k=(m+{1\over 2}){2\p\over L}$ and 
$k_0=(m_0+{1\over 2}){2\pi\over L}$ with $m,m_0=0,1,\ldots,L/a-1$.
To simplify the notations
we write $\int d\xx=a^2\sum_{\xx\in\L}$ and $\int d\kk={1\over L^2}\sum_{\kk\in\DD}$.

-)$\psi_\xx,\bar\psi_\xx$, $\xx\in\L$ are a finite set of
{\it Grassmann spinors} 
and $P_N(d\psi)$ is the fermionic integration 
with propagator
$$g(\xx-\yy)=\int d\kk {-i\not\pp+m\over \pp^2+m^2}
e^{-i\pp(\xx-\yy)}
\chi_{N}(\kk)\Eq(11.1)$$
where $\chi_{N}(\kk)$ is a {\it smooth cutoff function}
selecting momenta $|\kk|\le \g^N$ with $\g>1$ and $N$ a positive integer.
We assume $\g^{N}<<a^{-1}$,
that is the lattice cutoff is removed before the fermionic cutoff
(we are essentially considering a continuum model with a momentum
regularization).

-)The $\g$'s matrices are
$$\g^0=
 \pmatrix{0&1\cr
          1&0\cr}\;,
 \qquad
 \g^1=
 \pmatrix{0&-i\cr
          i&0\cr}\;,
 \qquad
 \g^5=-i\g^0\g^1
 =
 \pmatrix{1&0\cr
          0&-1\cr}\;.\Eq(gam)$$

-)$A_\xx=(A_{0,\xx},A_{1,\xx})$ are Euclidean boson fields 
with periodic boundary conditions and Gaussian measure $P(dA)$
with covariance
$$<A_{\m,\xx} A_{\n,\yy}>=v(\xx-\yy)\d_{\m,\n}=\d_{\m,\n}
\int d\pp e^{-i\pp(\xx-\yy)}v(\pp)\Eq(2)$$
\*
Integrating the bosonic variables $A$ one 
can rewrite \equ(1) as 
$$e^{\WW_{N,L}(J,\phi)}=
\int 
P_N(d\psi) 
e^{{1\over 4}\int d\xx d\yy v(\xx-\yy)
[e \bar\psi_{\xx}\g^\m\psi_{\xx}+J_{\m,\xx}][e \bar\psi_{\yy}\g^\m\psi_{\yy}+J_{\m,\yy}]+\int d\xx
[{\phi_\xx\bar\psi_\xx\over\sqrt{Z}}+
{\bar\phi_\xx\psi_\xx
\over \sqrt{Z}}]}\Eq(1a)$$
The {\it Schwinger functions} are defined by
$$<\prod_{i=1}^n \psi_{\xx_i}\prod_{i=1}^n\bar\psi_{\yy_i}
\prod_{i=1}^m j^\m_{\zz_i}>_{N,L}=
{\partial^{2n+m}\WW_{N,L}(J,\phi,\bar\phi)\over\partial \phi_{\xx_1}
...\partial \phi_{\xx_n}
\partial\bar\phi_{\yy_1}
...\partial \phi_{\yy_n}
\partial J_{\m_1,\zz_1}
...\partial J_{\m_n,\zz_m}}\Big|_{J=\phi=0}\Eq(14)$$
where $j^\m_{\xx}=\bar\psi_\xx\g^\m\psi_\xx$. Of course
the following trivial identities hold
$$e v(\pp)<j^{\m}_\pp\psi_\kk\bar\psi_{\kk+\pp}>=
<A_{\m,\pp}\psi_\kk\bar\psi_{\kk+\pp}>\Eq(al9002)$$
and                           
$$e v(\pp) <j^{5,\m}_\pp\psi_\kk\bar\psi_{\kk+\pp}>
=ie v(\pp) \e^{\m,\n} <j^{\n}_\pp\psi_\kk\bar\psi_{\kk+\pp}>
=i  \e_{\m,\n} <A_{\n,\pp}\psi_\kk\bar\psi_{\kk+\pp}>\Eq(al002)$$
where we have used that $j^{5,\m}=i\e^{\m,\n} j^{\n}$, $\e_{\m,\n}=-\e_{\n,\m}$,
$\e_{0,1}=-1$.

Depending 
on the explicit form of $v(\pp)$, to the functional integral 
\equ(1) correspond several models: if $v(\pp)=\pp^{-2}$ and $m=0$ it 
is a regularized version of the {\it Schwinger model}, if 
$m\not=0$ is a version of $QED_2$ in the Feynmann gauge,
if $v(\pp)=
(\pp^2+M^2)^{-1}$ it corresponds to the {\it Vector-gluon model}
of [GR]; an ultraviolet cutoff can be eventually imposed if necessary. 
Particularly interesting is the case
$v(\pp)=1$ (that is $v(\xx-\yy)=\d(\xx-\yy)$)
corresponding to the massive {\it Thirring model} (with a definite sign
of the interaction).

The Schwinger functions \equ(14) are well defined if the cutoffs (the volume $L$
and the momentum cutoff $N$)
are {\it finite}; the main problem is to show that, 
choosing properly the bare parameters $Z,m$ (eventually depending from the cutoffs)
$
<\prod_{i=1}^n \psi_{\xx_i}\prod_{i=1}^n\bar\psi_{\yy_i}\prod_{i=1}^m j^\m_{\zz_i}>_{N,L}$
has a well defined non trivial limit as $N,L\to\io$.

In this paper we will prove that, if the bosonic propagator
decays fast enough in momentum space and 
for small coupling,
the cutoffs $L,N$
can be removed in the Schwinger functions for any finite $m$ and $Z$.
We will start from the fermionic representation
\equ(1a) and the Grassmann functional integral is nonperturbatively
evaluated
by a multiscale analysis in which each step is proved to be well defined
by tree expansion methods and determinant bounds
(for a tutorial introduction to such techniques see [GM]); the 
massless
limit is controlled using the methods introduced
in [BM] allowing the implementation of WI 
(approximate,due to cutoffs) based on local Gauge invariance
at each integration step.

By performing the local gauge
transformation $\bar\psi_\xx
\to e^{\a_{\xx}}\bar\psi_{\xx}$, 
$\psi_\xx
\to e^{-\a_{\xx}}\psi_{\xx}$ or 
$\bar\psi_\xx
\to e^{\g_5\a_{\xx}}\bar\psi_{\xx}$, 
$\psi_\xx
\to e^{-\g_5\a_{\xx}}\psi_{\xx}$
in \equ(1) we get, in the case $m=0$
$$\eqalign{
&-i\pp_\m
\la j^\m_\pp\psi_\kk\bar\psi_{\kk-\pp}\ra
=\la\psi_{\kk-\pp}\bar\psi_{\kk-\pp}\ra
-\la\psi_{\kk}\bar\psi_{\kk}\ra+
\D_{N,L}^0(\kk,\kk-\pp)\cr
&-i\pp_\m\la j^{5,\m}_\pp\psi_\kk\bar\psi_{\kk-\pp}\ra
=\g^5[\la\psi_{\kk-\pp}\bar\psi_{\kk-\pp}\ra
-\la\psi_{\kk}\bar\psi_{\kk}\ra]+\D_{N,L}^5(\kk,\kk-\pp)
\;,}\Eq(90)$$
where
$$\eqalign{
&\D_{N,L}^0(\kk,\kk+\pp)=\int d\kk' 
C^{\m,N}_{\kk',\pp}<\bar\psi_{\kk'}\g^\m\psi_{\kk'+\pp}
\psi_\kk\bar\psi_{\kk+\pp}>_{N,L}\cr
&\D_{N,L}^5(\kk,\kk+\pp)=\int d\kk' 
C^{\m,N}_{\kk',\pp}<\bar\psi_{\kk'}\g^\m\g^5\psi_{\kk'+\pp}
\psi_\kk\bar\psi_{\kk+\pp}>_{N,L}
\;,}\Eq(al900)$$
with 
$$C^{\m,N}_{\kk,\pp}=([\chi_{N}(\kk)]^{-1}-1)\kk^\m-
([\chi_{N}(\kk-\pp)]^{-1}-1)(\kk^\m-\pp^\m)\Eq(92)$$
The last term in \equ(90) is due to the presence of 
the cutoff function breaking the formal Gauge invariance
of the action (it is formally vanishing if $\chi_{N}(\kk)=1$)
and it is the average of
the highly non local operator
$\int d\kk' 
C^{\m,N}_{\kk',\pp} \bar\psi_{\kk'}\g^\m\psi_{\kk'+\pp}$.
We prove the following result.
\vskip.3cm
{\rm THEOREM 1} {\it Let us consider the generating functional \equ(1) with $Z=1$,
$|v(\xx)|\le C$,
$\int d\xx [|v(\xx)|+|\partial v(\xx)|+|\xx| |v(\xx)|]
\le C$ for a suitable constant $C$ and
and $e$ small enough; then 
the Schwinger functions \equ(14) are such that the limit
$$\lim_{L,N\to\io}
\la \prod_{i=1}^n \psi_{\xx_i}\prod_{i=1}^n\bar\psi_{\xx_i}\prod_{i=1}^m j^\m_{\zz_i}\ra_{L,N}=
\la \prod_{i=1}^n \psi_{\xx_i}\prod_{i=1}^n\bar\psi_{\xx_i}\prod_{i=1}^m j^\m_{\zz_i}\ra\Eq(4000)$$
exists at noncoincinding points uniformly in the fermionic mass and is non trivial.

In the massless case $m=0$ the WI \equ(90) holds 
and 
$$\eqalign{
&\lim_{L,N\to\io} \D_{N,L}^0(\kk,\kk+\pp)=
-{e\over 4\pi}(-i\pp_\m) <A_{\m,\pp}\psi_\kk\bar\psi_{\kk+\pp}>\cr
&\lim_{L,N\to\io}\D_{N,L}^5(\kk,\kk+\pp)=
{e\over 4\pi}(-i\pp_\m )i \e^{\m,\n} <A_{\n,\pp}\psi_\kk\bar\psi_{\kk+\pp}>
\;,}
\Eq(4001)
$$
}
\*
The above result says that 
the correction terms $\D^0_{N,L},\D^5_{N,L}$ to the Ward Identities \equ(90), produced
by the presence of the cutoff functions, generate the anomalies when the cutoffs
are removed. 
Similar WI with any number of fermionic fields can be obtained
and this can be read as 
$$\partial_\m j_\m= -{e\over 4\pi}\partial_\m A_\m
\quad
\partial_\m j^5_\m={e\over 4\pi} i\e^{\m,\n}\partial_\m A_\n\Eq(ti)$$
that is the anomaly is non-renormalized by higher orders,
in agreement with the Adler-Bardeen theorem in $d=2$
[GR] based on a cancellation between an infinite collection
of Feynmann diagrams and with a momentum regularization for the fermionic loop.
The main point is however that Theorem 1 is a
{\it non perturbative} version of the Adler-Bardeen theorem, which is based
neither on a Feynmann graphs expansion
(for which convergence cannot be proved) nor on an exact
evaluation of the functional integrals, which is not possible without
approximations. The main technical tool is an expansion
in terms of product of determinants, which allow us to implement 
the cancellations among Feynmann graphs due to the relative signs
and it has good convergence properties. It turns out 
that all higher orders contributions
to the anomaly vanish removing the cut-offs, and this 
is proved partly expanding the determinants in such a way that 
the good convergence properties are not lost. 

From our construction an almost complete characterization
of the Schwinger function is also obtained; for instance we can prove that
the two point function
$<\psi_\xx\bar\psi_\yy>$ decays for large $|\xx-\yy|$ as $e^{-m^{1+\hat\h}|\xx-\yy|}|\xx-\yy|^{-1-\h}$
where $\h=a e^4+O(e^6)$ and $\hat\h=-b e^2+O(e^2)$ while for $\xx\to\yy$
it diverges as $|\xx-\yy|^{-1}$. The condition assumed in Theorem 1 for the bosonic propagator
are verified for instance by 
$$
v(\pp)=\int d\pp {e^{-i\pp(\xx-\yy)}\over \pp^2+M^2}\chi_K(\pp)\Eq(1zzz) 
$$
corresponding a massive boson propagator with an ultraviolet cut-off, 
which could be removed with some more
technical effort.
\vskip.5cm
\sub(2.1aklsx){\it Local interaction.} The previous result says
that the anomaly nonrenormalization holds if the bosonic
propagator in momentum space
decays fast enough and it is finite; the question then naturally rises
if the anomaly nonrenormalization is valid
also if the bosonic propagator does not decay at all,
as in the case of the Thirring model in which $v(\pp)$ is a constant.
In a companion paper [BFM] the case $v(\pp)=1$
has been studied, and it has been found that
the functional integral \equ(1) still defines
a set of Schwinger functions removing cutoffs, in the limit
$L,N\to\io$, that is \equ(4000) still holds
provided that we choose $Z=Z_N, m\equiv m_N$ 
depending on the ultraviolet cutoff, that is
$$
Z_N=\g^{-N\h}\big(1+{\rm O}(e^4)\big)
\quad\quad m_M=m \g^{-\bar\h N}\big(1+{\rm O}(e^2))\Eq(44)
$$
with $\h$ and $\bar\h$ independent of $m$ and such that
$\h=a e^4+{\rm O}(e^6)$, $\bar\h=b e^2+{\rm O}(e^4)$, $a,
b>0$. In the massless case $m=0$ the WI \equ(90) holds 
but \equ(4001) has to be replaced by
$$\eqalign{
&
\lim_{L,N\to\io} \D_{N,L}^0(\kk,\kk+\pp)=
[-{e\over 4\pi}+c_+e^3+F_\a]
\pp_\m <A_{\m,\pp}\psi_\kk\bar\psi_{\kk+\pp}>\cr
& \lim_{L,N\to\io} \D_{N,L}^0(\kk,\kk+\pp)\D_{N,L}^5(\kk,\kk+\pp)=
i[{e\over 4\pi}+c_+e^3+F_\a]\pp_\m \e^{\m,\n} <A_{\n,\pp}\psi_\kk\bar\psi_{\kk+\pp}>
\;,}
\Eq(40011)
$$
with $c_+>0$ {\it non vanishing} and $|F_\a|\le C e^5$.
This means that if $v(\pp)=1$ 
{\it the anomaly has higher orders corrections}, that is \equ(ti) has to be replaced by
$$
\partial_\m j_\m=i[-{e\over 4\pi}+c_+e^3+e F_+]\partial_\m A_\m\quad 
\partial_\m j^5_\m=[{e\over 4\pi}+c_+e^3+e F_]i\e^{\m,\n}\partial_\m A_\n\Eq(13ss)
$$
This result of course implies that 
{\it one cannot replace the determinant in \equ(12)
by a quadratic exponential}; the contribution
of the irregular fields is not negligible when $v(\pp)=1$.
In Appendix 3 an explicit second order verification of 
\equ(4001) \equ(13ss) has been included.
\equ(13ss) is apparently contrast with the Adler-Bardeen theorem [AB], but indeed this is not the case. 
In the [AB] analysis for $QED_4$, 
an ultraviolet cut-off $K$ has been introduced for the boson propagator,
and it is implicitly assumed that it is 
removed {\it after} the ultraviolet cut-off for the fermionic
propagator; 
moreover the bare parameters are chosen as a function
of $K$. In the model \equ(1), if $v(\pp)$ decays for large momenta
the theory is superrinormalizable, while if $v(\pp)=1$ is just
renomalizable like $QED_4$. Proceeding analogously to [AB] (in a non perturbative framework)
an ultraviolet cut-off can be introduced also in the bosonic propagator
for instance by replacing $v(\pp)=1$ with 
$e^{-\pp^2 K^{-2}}$.
There are then {\it two} ultraviolet cutoffs,
corresponding to the fermionic or bosonic
propagator, and {\it depending
which cutoff is removed first different anomalies are found} as functions of the bare parameters.
If $K$ is removed before the fermionic cutoff $N$, that is $N\to\io, K\to\io$,
we are essentially considering the case $v(\pp)=1$ discussed in [BFM];
it holds that the anomaly is given by \equ(13ss), that is it is non linear in $e$
but it is {\it renormalized} by higher orders.
On the other hand if the fermionic cutoff is removed
first a completely different result holds.
\*
{\rm THEOREM 2}
{\it Assume that $v(\pp)=e^{-\pp^2 K^{-2}}$; it is possible 
to find bare parameters $Z=Z_K,m=m_K$ such that the limit
$$\lim_{L\to\io}\lim_{K\to\io}\lim_{N\to\io} 
\la \prod_{i=1}^n \psi_{\xx_i}\prod_{i=1}^n\bar\psi_{\xx_i}
\prod_{i=1}^m j_\m(\zz_i)\ra_{L,N,K}\Eq(14dfd)$$
exists and it is non trivial, and \equ(90) holds togheter with \equ(4001), that is anomaly
nonrenormalization holds.}
\*
This means that the
the anomaly nonrenormalization holds if the fermionic cutoff is removed first,
while is violated is the bosonic cutoff is removed first.
The limit $N\to\io,K\to\io$ corresponds
to a fermionic functional integral \equ(1a) with a local
Thirring current-current interaction $j_\m(\xx)j_\m(\xx)$;
the opposite limit $K\to\io,N\to\io$
is similar to the one used in the original [AB] paper. 

In \S 2 we will describe our multiscale integration
procedure, and in \S 3 theorem 1 and 2 are proved. A second order
verification of our results is included for pedagogical reasons
in Appendix 3.
\*
\section(2,Multiscale Integration)
\vskip.5cm
\sub(1.1aklx){\it Multiscale analysis}

It is convenient to adopt Weyl notation.
Calling $\psi_\xx=(\psi_{+,\xx}^-, \psi_{-,\xx}^-)$, 
and  $\psi^\dagger_\xx=(\psi_{+,\xx}^+, \psi_{-,\xx}^+)$, 
$\bar\psi=\psi^\dagger_\xx\g_0$,
the {\it Generating Functional} \equ(1) can be written as
$$\eqalign{
 &e^{\WW_{N,L}(J,\phi)}
 =\int\!\!P(d\ps)\exp\Big\{\l {1\over 2}
 \sum_{\o=\pm}\int\!d\xx\ 
 \hp^{(\le N)+}_{\xx,\o}
\hp^{(\le N)-}_{\xx,\o}\hp^{(\le N)+}_{\xx,-\o}\hp^{(\le N)-}_{\xx,-\o}+\cr
&\int\!d\xx
J_{\xx,\o}\ps^{(\le N)+}_{\xx,\o}\ps^{(\le N)-}_{\xx,\o}
+\sum_\o\int\!d\xx\ 
\lft[\f^{+}_{\xx,\o}\ps^{(\le N)-}_{\xx,\o}+\ps^{(\le N)+}_{\xx,\o}\f^{-}_{\xx,\o}\rgt]
\Big\}\;,}\Eq(gf)$$
where  $\l=e^2$ and
$$P(d\ps^{(\le N)})\defi
\prod_{\kk\in\DD}
\prod_{\o=\pm}{\der\hp_{\kk,\o}^{(\le N)+}\der\hp_{\kk,\o}^{(\le N)-}
\over \NN_{N}(\kk)}
\exp\left\{-
 {1\over L^2}\sum_{\o,\o'=\pm}
 \sum_{\kk\in \DD}
 {T_{\o,\o'}(\kk)\over C^{-1}_{N}(\kk)}
 \hp^{(\le N)+}_{\kk,\o}\hp^{(\le N)-}_{\kk,\o'}\right\}
\Eq(distr)$$
and 
$$ T_{\o,\o'}(k)
 \defi
 \pmatrix{ D_+(\kk) & -m \cr
           -m & D_-(\kk)\cr}_{\o,\o'}\;;
 \qquad
 D_\o(\kk)\defi-ik_0+\o k_1\;.\Eq(mm)$$
and $\{J_{\xx,\o}\}_{\xx,\o}$ are commuting variables, while
$\{\f^\s_{\xx,\o}\}_{\xx,\o,\s}$ are anticommuting.
Finally, the normalization 
of the fermionic measure is
$\NN_{N}(\kk)\defi -(1/L^4)|\kk|^2 C^2_{N}(\kk)$.

The function $C_{N}^{-1}(\kk)$ is defined in the following way; 
$\chi_0\in C^\io(\RRR_+)$ is a non-negative, non-increasing function such that
$$\chi_0(t)\defi
\lft\{\matrix{
1\hfill&\hfill{\rm if\ }0\le t\le 1\cr 
0\hfill&\hfill{\rm if\ } t\ge \g_0\;,}\rgt.$$
for any choice of  $\g_0:1<\g_0\le\g$; 
and we define, for any $h$, 
$$f_j(\kk)\defi
\chi_0\lft(\g^{-j}|\kk|\rgt)-\chi_0\lft(\g^{-j+1}|\kk|\rgt)\Eq(bbb)$$
and $C_{N}^{-1}(\kk)=\sum_{j=-\io}^N f_j(\kk)$; hence 
$C_{N}^{-1}(\kk)$ acts
as a cutoff for momenta $|\kk|\ge \g^{N+1}$ (ultraviolet region).
By well known properties of Grassmann integrals (see for instance [GM])
we can write
$$P(d\ps^{(\le N)})=\prod_{h=-\io}^N P(d\ps^{(h)})\Eq(dis1)$$
where $P(d\ps^{(h)})$ is given by \equ(distr) with $f_h(\kk)$
replacing $C_{N}^{-1}(\kk)$. We integrate iteratively starting from the highests scales.
We define
$$\int d\xx |\xx| |v(\xx)|=\g^{-h_M}\Eq(gggn)$$ 
and the integration procedure is different 
for scales greater or smaller than $h_M$.
\vskip.5cm
\sub(1.gh){\it Ultraviolet integration}

We show inductively that, for any $h_M\le k\le N$ 
$$e^{- L^2 F_k}\int P
(d\psi^{(\le k)})e^{-\VV^{(k)} (\psi^{(\le k)},\phi,\hat J)
}\;,\Eq(2.17)$$
where the Grassmann integration $P(d\psi^{(\le k)})$ 
is equal to $P(d\psi^{(\le N)})$ with the cutoff function
$C_{N}(\kk)$ replaced by $C_{k}(\kk)$,
$$\VV^{(k)} (\psi^{(\le k)},\phi,J)=\bar\VV^{(k)} (\psi^{(\le k)},\hat J)+
\BB^{(k)}(\psi^{(\le k)},\phi, J)\Eq(pppkdd)$$
where
$$\bar\VV^{(k)}=\sum_{l,\underline\o,\underline\e}\int 
d\xx_1...d\xx_{2l} W^{(k)}_{2l,m,\underline\o,\underline\e}
\prod_{i=1}^{2l} \psi^{\e_i{\le k}}_{\xx_i,\o_i}
\prod_{i=1}^m J_{\o_i}(\xx_i)\Eq(2.17a)
$$
and 
$$\BB^{(k)}=\sum_\o\int\!d\xx\ 
\lft[\f^{+}_{\xx,\o}\ps^{(\le k)-}_{\xx,\o}+\ps^{(\le k)+}_{\xx,\o}\f^{-}_{\xx,\o}\rgt]
+\sum_{\bar m}\sum_{\underline\o,\underline\o',\underline\s}\int d\xxx$$
$${\partial^{\bar m} \bar\VV^{(k)}\over\partial\ps^{(\le k)\s_1}_{\xx_1,\o_1}...
\partial\ps^{(\le k)\s_m}_{\xx_m,\o_m}}g^{[k,N]}_{\o_1,\o'_1}(\xx_1-\yy_1)\phi^{\s_1}_{\yy_1,\o'_1}
...g^{[k,N]}_{\o_m,\o'_m}
(\xx_{\bar m}-\yy_{\bar m})\phi^{\s_{\bar m}}_{\yy_1,\o'_{\bar m}}+\int d\xx d\yy
g^{[k,N]}_{\o,\o'}(\xx-\yy)\phi^{+}_{\xx,\o}\phi^{-}_{\yy,\o'}
$$
where $g^{[k,N]}_{\o,\o'}(\xx-\yy)=\sum_{i=k}^N g^{(i)}(\xx-\yy)$.
In order
to inductively prove \equ(2.17)
for $\g^{h_M}\le k<N$ we proceed as follows. We
introduce the {\it localization operator} as a linear operator
acting on the kernels $W^{(k)}_{2l,m,\underline\o,\underline\e}$ in the following way:
$$\LL W^{(k)}_{2l,m}=
\cases{W^{(k)}_{2l,m}\hskip.7truecm {\rm if}
\hskip.7truecm l=1,2\quad\quad m=0\cr
W^{(k)}_{2l,m} \hskip.7truecm {\rm if}
\hskip.7truecm l=1\quad\quad m=1\cr
0\quad\quad\quad\quad\quad\quad\quad\quad\quad\quad\quad\quad\quad
{\rm otherwise} \hskip.7truecm
\;} \Eq(2.20)$$
We also define $\RR$ as $\RR=1-\LL$ and rewrite the r.h.s. of
\equ(2.17) as
$$e^{- L^2 F_k}\int P(d\psi^{(\le k)})e^{-\LL\VV^{(k)} (\psi^{(\le k)},\phi,J)
-\RR\VV^{(k)} (\psi^{(\le k)},\phi,J)}\;,\Eq(2.21)$$
where by definition $\LL\VV^{(k)}$ can be written as
$$\eqalign{&\LL\bar\VV^{(k)}=\sum_{\o}\int d\xx d\yy\, n_{k,\o}(\xx,\yy)\,
\psi^{(\le k)+}_{\xx,\o} \psi^{(\le k)-}_{\yy,\o}+
\sum_{\o,\o'}\int d\xx d\yy d\zz 
(1+Z^{(2)}_{k,\o',\o}(\zz;\xx,\yy))
J_{\o'}(\zz)\psi^{(\le k)+}_{\xx,\o} \psi^{(\le k)-}_{\yy,\o}+
\cr
&
-\sum_{\o,\o'} \int d\xx_1 d\xx_2 d\xx_3 d\xx_4\,
\l_{k,\o,\o'}(\xx_1,\xx_2,\xx_3,\xx_4)\,\psi^{(\le k)+}_{\xx_1,\o}
\psi^{(\le k)-}_{\xx_2,\o}\psi^{(\le k)+}_{\xx_3,\o'}
\psi^{(\le k)-}_{ \xx_4,\o'}\;.\cr}\Eq(2.22)$$
We write
$$\eqalign{&e^{- L^2\b (F_k+t_k)}\int P
(d\psi^{[\le k-1]})
\,\int P(d\psi^{(k)}) 
e^{-\LL\VV^{(k)} (\psi^{(\le k-1)}+\psi^{(k)},\phi,J)
-\RR \VV^{(k)} (\psi^{(\le k-1)}+\psi^{(k)},\phi,J)
} \;,\cr}\Eq(2.26)$$
with $P(d\psi^{(k)})$ a Grassmann Gaussian
integration with $f_k$ replacing $C_N^{-1}$,
and the corresponding propagator $g^{(k)}_{\o,\o'}(\xx,\yy)$
is bounded by,for any $N>1$
$$|g^{(k)}(\xx,\yy)|\le \g^k {C_{\bar N}\over 1+[\g^k|\xx-\yy|]^{\bar N}}\Eq(2.26aa)$$ 
If we now define
$$e^{-\VV^{(k-1)}(\psi^{(\le k-1)},\phi,J)- L^2\tilde F_k}=
\int P(d\psi^{(k)}) e^{-\LL\VV^{(k)} (\psi^{(\le k-1)}+\psi^{(k)},\phi,J)
-\RR \VV^{(k)} (\psi^{(\le k-1)}+\psi^{(k)},\phi,J)
}\;,\Eq(2.29)$$
it is easy to see that the procedure can be iterated.
In this way we have written the kernels $W^{(k)}_{2l,m}$
as functions of {\it running coupling functions}
$v_k(\underline\xx)=\l_k, n_k, Z^{2}_k$ with $k\ge h$;
the main advantage of this procedure 
is that the kernels $W^{(k)}_{2l,m}$
can be bounded 
if the running couplings are small enough.
Denoting by
$||f||={1\over L^2}\int \prod_{i=1}^n d\xx_i |f(\xx_1,..,\xx_n)|$
the kernels obey to the following dimensional bounds, see Appendix 1.
\*
{\bf Lemma 1}
{\it Assume that $||v_k||\le C\l$
for $k\ge h$, for a suitable constant $C$;
then it holds, if $\bar C$ is a constant 
$$||W^{(h)}_{2l,m}||\le \bar C\l
\g^{-h(l+m-2)}\Eq(2.36)$$}
\*
In order to use the above result to prove
that the kernels $W^{(h)}_{2l,m}$ are bounded, one has to show
that the running coupling functions are small.
By construction it holds that 
$$\eqalign{
\l_{k-1}(\xx_1,\xx_2,\xx_3,\xx_4)&=\l v(\xx_1-\xx_3)\d(\xx_1-\xx_2)
\d(\xx_3-\xx_4)+\sum_{h=k}^N W^{(h)}_{4,0}\quad
n_{k-1}(\xx_1,\xx_2)=\sum_{h=k}^N W^{(h)}_{2,0}(\xx_1,\xx_2)\cr
&
Z^{(2)}_k(\xx_1,\xx_2,\xx_3)=1+\sum_{h=k}^N W^{(h)}_{2,1}
(\xx_1,\xx_2,\xx_3)
;.\cr}\Eq(2.22aa)$$
The bound \equ(2.36) cannot be used in \equ(2.22aa)
to show that the running coupling constants are small,
and we have to improve it.
Defining
$$H_{n,m}^{(k)}=\sum_{h=k}^N W_{n,m}^{(h)}\Eq(marmat)$$
we prove the following result.
\*
{\bf Lemma 2}
{\it Assume that,for a suitable constant $C$ 
$$\sup_{j\ge k-1} ||\l_j(\underline\xx)||\le C\l\quad 
\sup_{j\ge k-1} ||n_j(\underline\xx)||\le C \l\quad
\sup_{j\ge k-1} ||Z^{(2)}_j(\xx)||\le C\l\Eq(ddc)$$
Then, for a suitable $C_1$
$$||H^{(k)}_{2,0}(\xx,\yy)||\le C_1 \g^{-k}\l
\quad ||H^{(k)}_{4,0}(\xx_1,..,\xx_4)||\le C_1 \g^{-k}\l^2\quad
||H^{(k)}_{2,1}(\xx_1,\xx_2,\xx_3)||\le C_1 \g^{-k}\l\Eq(kkkk)$$ }
\*
\*
{\it Proof.}
The proof is done by induction. First one proves \equ(kkkk)
for $k=N$ (\equ(ddc) is of course verified).
Moreover 
if \equ(kkkk)
is true for $j\ge k-1$, of course the running coupling constants 
are bounded; then it is sufficient to prove \equ(kkkk) for $j=k$.
The proof then is reduced to the verification of the bounds
\equ(kkkk) if \equ(ddc) are verified, and this is done below
distinguishing the different cases.
\vskip.5cm
\sub(1.1aklxa){\it Two fermionic lines}
\*
We start considering the massless case $m=0$.
We define the {\it truncated expectation}, if $X_i$ are momomials
in $\psi^{[k,N]}$, in the following way 
$$\EE^T_{k,N}(X_1;X_2;..;X_n)={\partial^n\over\partial\l_1...\partial\l_n}
\log \int P_{k,N}(d\psi)e^{\l_1 X_1+..\l_n X_n}|_{\l_1=\l_2=..=\l_n=0}\Eq(at)$$
where $P_{k,N}(d\psi)$ is given by \equ(distr) with $C_{k,N}$ 
replacing $C_N$. For semplicity of notations we also denotate
$$\EE^T_{k,N}(X_1X_2...X_n)\equiv\EE^T_{k,N}(X_1;X_2;...;X_n)\Eq(as1)$$
It holds that
$$H^{(k)}_{2,0}(\xx,\yy)=
{\partial^2\over\partial\phi^+_{\o,\xx}\partial\phi^-_{\o,\yy}}
\sum_{n=1}^\io {1\over n!}\EE^T_{k,N}(V(\psi+\phi)...V(\psi+\phi))\equiv \sum_{n=1}^\io {1\over n!}\
{\partial^2\over\partial\psi^-_{\o,\xx}\partial\psi^+_{\o,\yy}}
\EE^T_{k,N}(V(\psi)...V(\psi))
\Eq(abd)$$
We define 
$\tilde\psi(\xx)=\psi^+_{\xx,1}\psi^-_{\xx,1}$, 
$\tilde\psi(\yy)=\psi^+_{\yy,-1}\psi^-_{\yy,-1}$ and
$\tilde\psi(\xx\cup\yy)=\psi^+_{\xx,1}\psi^-_{\xx,1}
\psi^+_{\yy,-1}\psi^-_{\yy,-1}$. Hence 
$V$ in \equ(abd) is given by 
$\int d\xx d\yy \l v(\xx-\yy)\tilde\psi(\xx\cup\yy)$
and we can write
$${\partial^2\over\partial\psi^+_\xx\partial\psi^-_\yy}
\EE^T_{k,N}(V(\psi)...V(\psi))=\Eq(ov1)$$
$$n{\partial^*\over\partial\psi^-_\yy}
\int d\tilde\yy \l v(\xx-\tilde\yy)
\EE^T_{k,N}(\psi^-_\xx
\tilde\psi(\tilde\yy);V;...;V)+
n\d(\xx-\yy)\int d\tilde\yy v(\xx-\tilde\yy)
\EE^T_{k,N}(\tilde\psi(\tilde\yy);V;...;V(\psi))\Eq(la1)$$
where ${\partial^*\over\partial\psi^-_\yy}$ means that 
the derivative cannot be applied over $\psi^-_\xx$;
in \equ(la1) we have separated the case in which the two external
lines are connected to the same coordinate from the case in which are connected
to different coordinates.

We use the following property
of truncated expectations, 
see for instance [Le], if 
$\tilde\psi(P_1\cup P_2)=[\prod_{i\in P_1}\psi^{\e_i}_{\xx_i}]
[\prod_{i\in P_2}\psi^{\e_i}_{\xx_i}]$
$$\EE^T(\tilde\psi(P_1\cup P_2)
\tilde\psi(P_3)...\tilde\psi(P_n))=\Eq(ttt1)$$
$$\sum_{K_1,K_2, K_1\cap K_2=0\atop K_1\cup K_2=3,..,n}(-1)^\pi
\EE^T(\tilde\psi(P_1)\prod_{j\in K_1}\tilde\psi(P_j))
\EE^T(\tilde\psi(P_2)\prod_{j\in K_2}\tilde\psi(P_j))+
\EE^T(\tilde\psi(P_1)
\tilde\psi(P_2)...\tilde\psi(P_n))\Eq(ttt1)$$
and $(-1)^\pi$ is the parity of the permutation
necessary to bring the Grassmann variables
on the r.h.s. of \equ(ttt1) to the original order.
Note that the number of terms in the sum in the r.h.s.
of \equ(ttt1) is bounded by $C^n$ for a suitable constant $C$.
Note that the same property holds if we replace $\psi$ with $\psi+\phi$
where $\phi$ is an external line.

By using \equ(ttt1) for the first addend of \equ(la1) we get, $V(j)=\int d\xx_j d\yy_j
\l v(\xx_j-\yy_j)\tilde\psi(\xx_j\bigcup\yy_j)$
$$
\int d\tilde \yy v(\xx-\tilde\yy)
{\partial^*\over\partial\psi^-_\yy}
\EE^T_{k,N}(\psi^-_\xx;
\tilde\psi(\yy);V(1);...;V(n))
+$$
$$\sum_{K_1,K_2, K_1\cap K_2=0\atop K_1\cup K_2=1,..,n}(-1)^\pi
\int d\tilde\yy v(\xx-\tilde\yy) 
[{\partial^*\over\partial\psi^-_\yy} 
\EE^T_{k,N}(\psi^-_\xx \prod_{j\in K_1}V(j))]
\EE^T_{k,N}(\tilde\psi(\yy)\prod_{j\in K_2} V(j))$$
where have used that the derivative 
applied on the second truncated expectation gives zero (it is the expectation
of an odd number of fields). 

If we define
$$<A_1;...;A_n>_T={\partial^n\over\partial\l_1...\partial\l_n}\log \int P_{k,N}(d\psi)
e^{-V+\sum_{i=1}^n\l_i A_i}|_{\underline\l=0}\Eq(def)$$
by summing over $n$ we get
$$H^{(k)}_{2,0}(\xx,\yy)=\int d\tilde\yy  \l v(\xx-\tilde\yy) 
<\tilde\psi(\tilde\yy)> {\partial^*\over\partial\psi^-_\yy} <\psi_\xx>+$$
$$\int d\tilde\yy\l v(\xx-\tilde\yy) 
{\partial^*\over\partial\psi^-_\yy} <\psi^-_\xx;
\tilde\psi(\tilde\yy)>_T+
 \d(\xx-\yy)\int d\tilde\yy v(\xx-\tilde\yy)
<\tilde\psi(\tilde\yy)>
\Eq(111)$$
\*
\insertplot{300pt}{100pt}%
{\ins{130pt}{60pt}{$+$}
\ins{50pt}{60pt}{$=$}
\ins{280pt}{60pt}{$+$}
}%
{verticiT11}{\eqg(1v)}

\* \vbox{{\ }

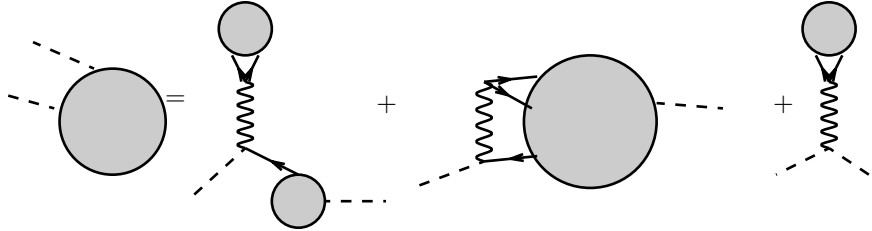
\centerline{Fig 1: Graphical representation of \equ(111)}
}
\*
By a multiscale integration similar to the one in the previous section we get,
see Appendix 1
$$||{\partial^*\over\partial\psi^-_\yy} <\psi^-_\xx;
\tilde\psi(\tilde\yy)>_T
||\le C \l \g^{-k}\Eq(2.36a)$$
Hence we get for the second addend in \equ(111)
the bound , using that $|v(\xx)|\le C$
$${1\over L^2}\int d\xx d\yy d\tilde\yy |\l v(\xx-\tilde\yy) 
{\partial^*\over\partial\psi^-_\yy} <\psi^-_\xx;
\tilde\psi(\tilde\yy)>_T |\le C\l ||
{\partial^*\over\partial\psi^-_\yy} <\psi^-_\xx;
\tilde\psi(\tilde\yy)>_T||\le C\l^2 \g^{-k}\Eq(vvbj)$$
On the other hand the first and third term in \equ(111)
are vanishing in the massless case. In fact
$$<\tilde\psi(\yy)>=0\Eq(11s)$$
as by translation invariance 
$<\tilde\psi(\yy)>=<\tilde\psi(0)>$.
As there are only diagonal propagators we note that by 
it is given by the integral of $(4n+2)/2$ diagonal propagators
and it is indipendent from $\tilde\yy$,
so it is vanishing by parity. Hence the integral of the first and last addend
in \equ(111) is vanishing.

%
%
%
%
%
\vskip.5cm
\sub(1.1aklxa){\it Two fermionic lines and one density line}
\*
We have to bound
$$H^{(k)}_{2,1}(\zz;\xx,\yy)=
\sum_{n=1}^\io {1\over n!}
{\partial^2\over\partial\psi^+_\xx\partial\psi^-_\yy}
\EE^T_{k,N}(\tilde\psi(\zz)V...V)\Eq(200.11)$$
and 
$${\partial^2\over\partial\psi^+_\xx\partial\psi^-_\yy}
\EE^T_{k,N}(\tilde\psi(\zz)V...V)=
n{\partial^*\over\partial\psi^-_\yy}
\int d\tilde\yy \l v(\xx-\tilde\yy)
\EE^T_{k,N}(\psi^-_\xx
\tilde\psi(\tilde\yy);\tilde\psi(\zz);V;...;V)+\Eq(200.1)$$
$$n\l\d(\xx-\yy)\int d\tilde\yy v(\xx-\tilde\yy)
\EE^T_{k,N}(\tilde\psi(\tilde\yy)\tilde\psi(\zz)...V(\psi))$$
where again ${\partial^*\over\partial\psi^-_\yy}$ means that 
the derivative cannot be applied over $\psi^-_\xx$;
that is we have distinguished the case 
the two external lines comes put from different
points or the same point.
The first addend can be written,by \equ(ttt1) as 
$$\int d\tilde\yy v(\xx-\tilde\yy)
{\partial^*\over\partial\psi^-_\yy}
\EE^T_{k,N}(\psi^-_\xx 
\tilde\psi(\tilde\yy);\tilde\psi(\zz);V;...;V)=
\int d\tilde \yy v(\xx-\tilde\yy)
{\partial^*\over\partial\psi^-_\yy}
\EE^T_{k,N}(\psi^-_\xx;
\tilde\psi(\tilde\yy);\tilde\psi(\zz);V(j);...;V(n)])
+$$
$$\sum_{K_1,K_2, K_1\cap K_2=0\atop K_1\cup K_2=1,..,n}(-1)^\pi
\int d\tilde\yy v(\xx-\tilde\yy) 
[{\partial^*\over\partial\psi(\yy)} \EE^T_{k,N}(\psi^-_\xx \prod_{j\in K_1}V(j))]
\EE^T_{k,N}(\tilde\psi(\zz)\tilde\psi(\tilde\yy)\prod_{j\in K_2}V(j))$$
$$\sum_{K_1,K_2, K_1\cap K_2=0\atop K_1\cup K_2=3,..,n}(-1)^\pi
\int d\tilde\yy v(\xx-\tilde\yy) 
[{\partial^*\over\partial\psi^-_\yy} \EE^T_{k,N}(\psi^-_\xx; \tilde\psi(\zz)
\prod_{j\in K_1}V(j))]
\EE^T_{k,N}(\tilde\psi(\tilde\yy)\prod_{j\in K_2}V(j))$$
We finally obtain, by summing over $n$
$$H^{(k)}_{2,1}(\zz;\xx,\yy)=\int d\tilde\yy d\tilde\xx \l v(\xx-\tilde\yy) 
<\tilde\psi(\tilde\yy)>{\partial^*\over\partial\psi^-_\yy}
<\psi^-_\xx;\tilde\psi(\zz)>_T+
\int d\tilde\yy  \l v(\xx-\tilde\yy) {\partial^*\over\partial\psi^-_\yy}
<\psi^-_\xx;\tilde\psi(\tilde\yy);\tilde\psi(\zz)>_T$$
$$+\int d\tilde\yy  \l v(\xx-\tilde\yy) 
<\tilde\psi(\tilde\yy);\tilde\psi(\zz)>_T
{\partial^*\over\partial\psi^-_\yy} <\psi^+_\xx>+
\d(\xx-\yy)\int d\tilde\yy v(\xx-\tilde\yy)
<\tilde\psi(\zz);\tilde\psi(\tilde\yy)>_T\Eq(111a)$$
\*
\*
\insertplot{300pt}{100pt}%
{\ins{130pt}{50pt}{$+$}
\ins{60pt}{50pt}{$=$}
\ins{265pt}{50pt}{$+$}
\ins{320pt}{50pt}{$+$}
}%
{verticiT14}{\eqg(1v)}

\* \vbox{{\ }

\centerline{Fig 2: Graphical representation of \equ(111a)}
}
\*
Again the first addend is vanishing in the massless case.
The integral of the second addend is bounded by, as shown in the Appendix
$${1\over L^2} \int d\xx  d\yy d\zz d\tilde\yy  
|\l v(\xx-\tilde\yy) {\partial^*\over\partial\psi^-_\yy}
<\psi^-_\xx;\tilde\psi(\tilde\yy);\tilde\psi(\zz)>_T|\le C 
||{\partial^*\over\partial\psi^-_\yy}
<\psi^-_\xx;\tilde\psi(\tilde\yy);\tilde\psi(\zz)>_T||\le C\l\g^{-2 k}
\Eq(lau1)$$
In the third and fourth addend appear a density-density
term which will be bounded in the following section
by $C\g^{-k}$ so that they are also $O(\l\g^{-k})$.
\vskip.5cm
\sub(1.1akljxa){\it Two density lines}
\*
We have to bound
$$H^{(k)}_{0,2}(\bar\zz,\yy)=\int d\xx \l v(\bar\zz-\xx)
\sum_{n=0}^\io {1\over n!}
\EE^T_{k,N}(\tilde\psi(\xx)\tilde\psi(\yy) V...V)$$
We can distinguish the case in which 
the two fields $\tilde\psi(\xx)$ are contracted
in the same point or not, so that
$$H^{(k)}_{0,2}(\bar\zz,\yy)=\int d\xx d\zz d\zz' \l v(\bar\zz-\xx) 
g^{[k,N]}_{\o,\o}(\xx-\zz)g^{[k,N]}_{\o,\o}(\xx-\zz')
{\partial^*\over\partial\psi^-_\zz}{\partial\over\partial\psi^+_{\zz'}}<\tilde\psi(\yy)>=$$
$$\int d\xx 
d\zz \l v(\bar\zz-\xx) [g^{(k,N)}_{\o,\o}(\xx-\zz)]^2
[\d(\zz-\yy)+\int d\zz' \l v(\zz-\zz')
 <\tilde\psi(\zz');\tilde\psi(\yy)>_T]+$$
$$\int d\xx d\zz d\zz' d\zz'' \l v(\bar\zz-\xx) 
g^{(k,N)}_{\o,\o}(\xx-\zz)g^{(k,N)}_{\o,\o}(\xx-\zz')
v(\zz'-\zz'')
{\partial^*\over\partial\psi^-_\zz}<\psi^-_{\zz'}
\tilde\psi(\zz'');\tilde\psi(\yy)>\Eq(pov)$$
where in the second line \equ(200.1) has been used.
\*
\insertplot{300pt}{100pt}%
{\ins{68pt}{55pt}{$=$}
\ins{125pt}{77pt}{$>$}
\ins{128pt}{40pt}{$<$}
\ins{250pt}{60pt}{$+$}
}%
{verticiT16}{\eqg(1v)}

\* \vbox{{\ }

\centerline{Fig 3: Graphical representation of \equ(pov)}
}
\*
The first addend of \equ(pov) can be rewritten as
$$\int d\xx 
d\zz v(\bar\zz-\zz)][g^{(k,N)}_{\o,\o}(\xx-\zz)]^2
[\d(\zz-\yy)+\int d\zz' \l v(\zz-\zz')
 <\tilde\psi(\zz');\tilde\psi(\yy)>_T]+$$
$$\int d\xx 
d\zz [v(\bar\zz-\xx)-v(\bar\zz-\zz)][g^{(k,N)}_{\o,\o}(\xx-\zz)]^2
[\d(\zz-\yy)+\int d\zz'\l v(\zz-\zz')
 <\tilde\psi(\zz');\tilde\psi(\yy)>_T]\Eq(pov1)$$
The first line of \equ(pov1) is vanishing in the massless case 
$$\int d\xx g^{(k,N)}_{\o,\o}(\xx-\zz)=0\Eq(llllm)$$
by the symmetry $g_{\o,\o}^{(k,N)}(r,r_0)=i \o g^{(k,N)}_{\o,\o}(r_0,-r)$;
on the other hand the second line can be written as
$$\int d\zz d\xx \l \int_0^1 dt [\partial_t v(\bar\zz-\zz+t(\zz-\xx)] 
(\xx-\zz)[g^{[k,N]}_{\o,\o}(\xx-\zz)]^2
[\d(\zz-\yy)+\int d\zz'\l v(\zz-\zz')
<\tilde\psi(\zz');\tilde\psi(\yy)>_T]
\Eq(mmm7)$$
We
perform the change of variables
$\rr_1=\xx-\zz\quad \rr_2=\bar\zz-\zz+t(\zz-\xx)$
with Jacobian $-1$, so that we can bound \equ(mmm7) as,using that $\int d\xx|\partial v(\xx)|\le C$
$$\eqalign{
&||
\int d\rr_1 d\rr_2 
\l \partial v(\rr_2) 
|\rr_1|g^{[k,N]}_{\o,\o}(\rr_1)]^2|| \le
C\l\sum_{k\le h_1\le h_2\le N}^N \g^{-3 h_2}
\g^{h_1}\g^{h_2}
\le C\l\g^{- k}\cr
&||
\int d\rr_1 d\rr_2 
\l \partial v(\rr_2) 
|\rr_1|g^{[k,N]}_{\o,\o}(\rr_1)]^2 \int d\zz' \l v(\xx-\rr_1-\zz')
<\tilde\psi(\zz');\tilde\psi(\yy)>_T||\le C\l\g^{- k}||H_{0,2}^k||\cr}\Eq(ccdaa)
$$
where we have used that $g^{[k,N]}(\rr)=\sum_{h=k}^N g^h(\rr)$.
We consider now the second addend of \equ(pov);
by \equ(111a) we get
$$\eqalign{
&\int d\xx \l v(\bar\zz-\xx) \int  d\zz d\zz' d\zz''  
v(\zz'-\zz'') g^{[k,N]}(\xx-\zz') g^{[k,N]}(\xx-\zz) 
{\partial^*\over\partial\psi_\zz}
<\psi^-_{\zz'};
\tilde\psi(\zz'');\tilde\psi(\yy)>_T
+\cr
&\int d\xx\l v(\bar\zz-\xx)
\int d\zz d\zz' d\zz'' d\zz'''
g^{[k,N]}_{\o,\o}(\xx-\zz) g^{[k,N]}_{\o,\o}(\xx-\zz')
H_{2,0}(\zz',\zz'')g^{[k,N]}_{\o,\o}(\zz''-\zz)]
v(\zz-\zz''')<\tilde\psi(\zz''');\tilde\psi(\yy)>_T\cr}\Eq(111b)
$$
\*
\*
\insertplot{300pt}{100pt}%
{\ins{130pt}{60pt}{$+$}
\ins{270pt}{60pt}{$+$}
}%
{verticiT15}{\eqg(1v)}

\* \vbox{{\ }

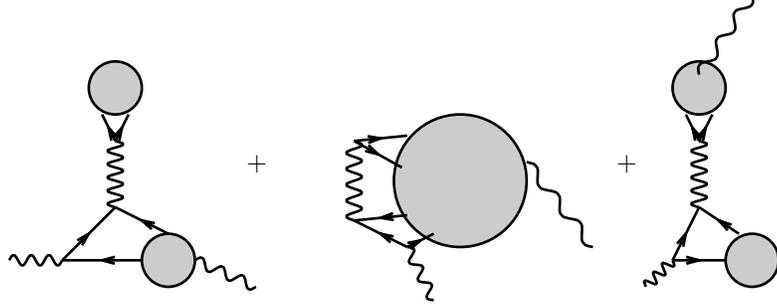
\centerline{Fig 4: Graphical representation of \equ(111b)}
}
\*
We prove in Appendix 1 that the first addend in \equ(111b) is bounded 
by 
$$||g^{[k,N]}(\xx-\zz') g^{[k,N]}(\xx-\zz) 
{\partial^*\over\partial\psi_\zz}
<\psi^-_{\zz'};
\tilde\psi(\zz'');\tilde\psi(\yy)>||
\le C\l\g^{-2k}\Eq(2.36b)$$
In order to bound the second addend 
$$\int d\zz d\zz' d\zz'' g_{\o,\o}^{[k,N]}(\xx-\zz')
H^{(k)}_{2,0}(\zz',\zz'')g^{[k,N]}_{\o,\o}(\zz''-\zz)
g^{[k,N]}_{\o,\o}(\xx-\zz)\int d\zz''' \l v(\zz-\zz''') 
<\tilde\psi(\zz''');\tilde\psi(\yy)>\Eq(alala)$$
which we can rewrite as, using the compact support properties of the propagator 
$$\sum_{k\le h_1, h_2\le N} \int d\zz d\zz' d\zz'' g_{\o,\o}^{(h_1)}(\xx-\zz')
H_{2,0}^{(k)}(\zz',\zz'')g^{(h_1)}_{\o,\o}(\zz''-\zz)
g^{(h_2)}_{\o,\o}(\xx-\zz)\int d\zz''' \l v(\zz-\zz''') 
<\tilde\psi(\zz''');\tilde\psi(\yy)>\Eq(alala1)$$
We distinguish now the case $h_1\le h_2$ or $h_1\ge h_2$; if 
$h_1\le h_2$ we integrate over $g^{(h_2)}_{\o,\o}$ and we use that $||H_{2,0}||\le 
C\l\g^{-k}$ and $|g^{(h_1)}_{\o,\o}|\le C\g^{h_1}$ so that we get 
$\sum_{k\le h_1\le h_2\le N}C\l\g^{-k}\g^{h_1}\g^{-h_1}\g^{-h_2}\le \bar C\l\g^{-k}$.
If $h_2\le h_1$ we use that $|g^{(h_2)}_{\o,\o}|\le C\g^{h_2}$ so that we get 
$\sum_{k\le h_2\le h_1\le N}C\l\g^{-k}\g^{-2 h_1}\g^{h_2}\le \bar C\l\g^{-k}$.
In both case we can bound \equ(alala) by
$C\g^{- k}||H^{(k)}_{0,2}||$.

By collecting all bounds we have found we have
$$||H_{0,2}^{(k)}||\le C_1\l\g^{- k}-C_2\l\g^{-k} ||H_{0,2}||$$
from which
$$||H_{0,2}^{(k)}||\le {C_1\l\g^{-k}\over 1+C_2\l\g^{-k}}\le C_3\l\g^{-k}
\Eq(fin)$$
\vskip.5cm
\sub(1.1akljxa){\it Four external lines}
\*
In this case we can write
$$H^k_{4,0}(\xx_1,\xx_2,\xx_3,\xx_4)=
\d(\xx_1-\xx_2)\d(\xx_3-\xx_4)
\l v(\xx_3-\zz')
H^k_{0,2}(\xx_1,\zz')+$$
$$\d(\xx_1-\xx_2)\l v(\xx_1-\zz)H^k_{1,2}(\zz;\xx_3,\xx_4)+
\bar H^k_{4,0}(\xx_1,\xx_2,\xx_3,\xx_4)$$
where the first two terms were evaluated before and the 
last term correspond to the four external lines attached at different
points; proceeding as above it can be written as
(in the massless case for semplicity)
$$\bar H^k_{4,0}=\int \l d\zz v(\xx_1-\zz)
{\partial^3\over\partial\psi_{\xx_2}\partial\psi_{\xx_3}\partial\psi_{\xx_4}}
<\psi^-_{\xx_1};\tilde\psi(\zz)>_T+
\int \l d\zz v(\xx_1-\zz)
{\partial\over\partial\psi_{\xx_2}}<\psi^-_{\xx_1}>_T
{\partial^2\over\partial\psi_{\xx_3}\partial\psi_{\xx_4}
}<\tilde\psi(\zz)>_T\Eq(llll11)$$
\*
\*
\insertplot{300pt}{100pt}%
{\ins{130pt}{60pt}{$+$}
\ins{50pt}{60pt}{$=$}
\ins{280pt}{60pt}{$+$}
}%
{verticiT10}{\eqg(1v)}

\* \vbox{{\ }

\centerline{Fig 5: Graphical representation of \equ(llll11)}
}
\*
As it is proved in the Appendix 1 
$$||{\partial^3\over\partial\psi_{\xx_2}\partial\psi_{\xx_3}\partial\psi_{\xx_4}}
<\psi^-_{\xx_1};\tilde\psi(\zz)>_T||\le C\l\g^{-k}\Eq(2.35g)$$
Finally the norm of last term in \equ(llll11) is bounded by
$$||{\partial\over\partial\psi_{\xx_2}}<\psi^-_{\xx_1}>_T||
||
{\partial^2\over\partial\psi_{\xx_3}\partial\psi_{\xx_4}
}<\tilde\psi(\zz)>_T||\le C\l \g^{-k}\Eq(2.35gg)$$
as follows from the previous bounds. 
\*
We have then proved \equ(kkkk)
in the massless case; in order to prove the bounds 
\equ(kkkk) in the massive case we note 
that we can write $g_{\o,\o}^k(\xx,\yy)=\bar g_{\o,\o}^k(\xx,\yy)+r_{\o,\o}^k(\xx,\yy)$
where $\bar g_{\o,\o}^k(\xx,\yy)$ is the propagator 
with $m=0$ and $r_{\o,\o}^k(\xx,\yy)$ verifies the bound \equ(2.26aa)
with an extra 
$[{m\over\g^k}]^2$; moreover $g_{\o,-\o}^k(\xx,\yy)=\bar g_{\o,-\o}^k(\xx,\yy)+
r_{\o,-\o}^k(\xx,\yy)$ where $\bar g_{\o,-\o}^k(\xx,\yy)$
is the Fourier transform of $[{m\over\g^k}]\bar g_{\o,\o}^k(\kk)
\bar g_{-\o,-\o}^k(\kk)$ and $r_{\o,-\o}^k(\xx,\yy)$
verifies the bound \equ(2.26aa)
with an extra 
$[{m\over\g^k}]^3$. Using the multilinearity of the determinants
appearing in the fermionic expectations, we can separate the contribution
in which all the propagators are massless (corresponding
to the cases treated above) plus a rest, with an extra 
$[{m\over\g^k}]$ with respect to the dimensional bound \equ(2.36).
The only case in which such improvment is not sufficient to get
\equ(kkkk) is for $H_{2,0}^{(k)}$. However when the external line
has the same $\o$ index, there is surely at least or a propagator 
$r_{\o,\o}^k(\xx,\yy)$ or a nondiagonal propagator $g_{\o,-\o}^k(\xx,\yy)$,
so that the improvment with respect to \equ(2.36) is given by 
$[{m\over\g^k}]^2$ and the bound \equ(kkkk) holds. The only
remaining case is when the two external lines have different
$\o$ index and all the propagators are massless (if they are not
there is an extra $[{m\over\g^k}]$) and 
there is only a nondiagonal propagator $\bar g_{\o,-\o}^k(\xx,\yy)$;
this case is then identical to the one trated in \S 2.4.
\vskip.5cm
\sub(1.aklxa){\it The infrared integration}
\*
After the integration of the scales $N,N-1,..h_M$
we get a functional integral of the form
$$e^{- L^2 F_{h_M}}\int P
(d\psi^{(\le h_M)})e^{-\VV^{(h_M)} (\psi^{(\le h_M)},\phi,J)}\;,\Eq(2.17a)$$
where $\VV^{(h_M)}$ given by
\equ(2.17a). The multiscale analysis of \equ(2.17a) has been
done in [BM] in all details and we will not repeat it here;
it turns out that after the integration of $h_M,...h$
one gets 
$$e^{- L^2 F_{h}}\int P_{Z_h,m_h}
(d\psi^{(\le h)})e^{-\VV^{(h)} (\sqrt{Z_h}\psi^{(\le h)},\phi,J)}\;,\Eq(2.17aa)$$
where $P_{Z_h,m_h}
(d\psi^{(\le h)})$ is given by \equ(distr) with $C_N$ replaced by $C_h$, 
wave function renormalization $Z_h$ and mass $m_h$;
moreover the effective interaction 
$\VV^{(h)}$ is $\l_h\int d\xx \psi^{(\le h)+}_{\o,\xx}
\psi^{(\le h)-}_{\o,\xx}\psi^{(\le h)+}_{-\o,\xx}\psi^{(\le h)-}_{\o,\xx}$ 
plus monomials in $\psi$ of higher orders.
As a consequence of remarkable cancellations due to the implementation
of Ward identites based a local phase transformation
at each iteration step, the effective counpling $\l_h$ remains close to 
its initial value
$$\l_h=\l+O(\l^2)\Eq(vfvhh)$$
and 
$$Z_h=\g^{-\h h}(1+O(\l))\quad\quad\m_h=\g^{-\tilde \h h}m(1+O(\l))\Eq(rcc100)$$
with $\h=a\l^2+O(\l^3)$ and $\tilde\h=b \l+O(\l^2)$
with $a,b$ positive constants. 
\equ(2.17aa) is found by a procedure similar to the previous one in
which the $\LL$ operation consists in computing the kernel $W^h(\kk)$
at zero momentum or, in coordinate space, it consist of computing
the external fields in the same coordinate point. Then, see [BM],
to each kernel $W^h$ is applied
$1-\LL$ which produces a derivative aplied on the external
fields, giving an extra $\g^h$, and a factor $(\xx-\yy)$,
if $\xx,\yy$ are the coordinate of the external fields,
wich can be bounded using that
by $\int d\zz |\zz| |g^i(\zz)|\le C \g^{-2i}$
or $\int d\zz |\zz| |v(\zz)|\le \g^{-h_M}\le \g^{-i}$.
In this way an expansion
for the Schwinger functions well defined 
in the limit $L,N\to\io$ is obtained.
\*
%
%
%
%
\*
\*
\section(2c,Ward Identities)
\vskip.5cm
\sub(1.34aklxa){\it The anomaly}
Performing 
the phase and chiral transformation 
$\psi^\pm_{\xx,\o}\to e^{\pm \a_{\xx,\o}}
\psi^\pm_{\xx,\o}$ in \equ(gf) and making derivatives
with respect to the external fields we get,if $\r_{\o,\xx}=\psi^+_{\o,\xx}\psi^-_{\o,\xx}$
$$D_{\o'}(\pp) \la\hat\r_{\o',\pp}\hp^-_{\o,\kk}\hp^+_{\o,\kk-\pp}\ra
=
\d_{\o,\o'}\left[\la\hp^-_{\o,\kk}\hp^+_{\o,\kk}\ra-
\la\hp^-_{\o,\kk-\pp}\hp^+_{\o,\kk-\pp}\ra\right]+
\D^{2,1,N}_{\o',\o}(\pp;\kk)\Eq(2.1)$$
where $\D^{2,1,N}_{\o',\o}(\pp;\kk)$ is the Fourier transform of
$$\D^{2,1,N}_{\o',\o}(\xx;\yy,\zz)\defi
\la\psi^-_{\yy,\o};\psi^+_{\zz,\o};\d T_{\xx,\o'}\ra_{L,N}\;,\Eq(2.2)$$
where
$$\d T_{\xx,\o}\defi
{1\over L^4}\sum_{\kk^+,\kk^-\atop\kk^+\not=\kk^-}
 e^{i(\kk^+-\kk^-)\xx} C^\e_{N;\o}(\kk^+,\kk^-)
\hp^+_{\kk^+,\o}\hp^-_{\kk^-,\o}\;,\Eq(int)$$
and 
$$C_{N;\o}(\kk^+,\kk^-)=[C_{N}(\kk^-)-1]D_\o(\kk^-)
-[C_{N}(\kk^+)-1]D_\o(\kk^+)\;.
\Eq(ccc)$$
We write
$$\D^{2,1,N}_{\o,\o'}(\pp;\kk)=v(\pp)\bar\n_N(\pp)
D_{-\o}(\pp) 
G_{-\o,\o'}^{2,1,N}(\pp;\kk)+\pp_i
R^{2,1,N}_{\o,\o',i}(\pp;\kk)\Eq(400.2)$$
where
$$\bar\n(\pp)=\l
\int {d\kk\over (2\pi)^2} {C_{\o,N}(\kk,\kk-\pp)\over D_{-\o}(\pp)}
g^{(\le N)}_\o(\kk)g^{(\le N)}_\o(\kk-\pp)\Eq(fhnk)$$
Note that if $\pp$ is fixed, in the limit $N\to\io$ we get, by changing variables $\kk\to \g^N\kk$
and expanding in $\g^{-N}\pp$, $\bar\n(\pp)=\n+O(\pp\g^{-N})$,
where $\n=\bar\n(\pp)|_{\pp=0}$
and, using some cancellations due
to the symmetry $g_\o(\kk)=-i\o g_\o(\kk^*)$ with $\kk^*=(k,k_0)$,
it holds that 
$$\bar\n=\l\int {d\kk\over (2\pi)^2} {k_0 \over |\kk|}
\chi'_0(|\kk|)D^{-1}_\o(\kk)={\l\over 4\pi}\int_0^\io d\r
\chi'_0(\r)=-{\l\over 4\pi}
\Eq(fhnk1)$$
We can obtain $\pp_i
R^{2,1,N}_{\o,\o',i}$ 
from the generating function
$$e^{\WW_{\D} (J,\hat J,\phi)}=
\int P(d\psi)e^{-V(\psi)+\sum_\o \int d\zz 
J(\zz)\psi^+_{\zz,\o}
\psi^-_{\zz,\o}+\sum_\o\int d\zz
[\psi^+_{\zz,\o}\phi^-_{\zz,\o}+\phi^+_{\zz,\o}\psi^-_{\zz,\o}]+
T_0(\hat J,\psi)-\n_{-,N}T_{-}(\hat J,\psi)}\Eq(fgbnn)$$
with
$$\eqalign{
&T_0(\hat J,\psi)={1\over L^4}\sum_{\kk,\pp} \hat J(\pp)
C_{N,\o}(\kk,\kk-\pp) 
\psi^+_{\kk-\pp,\o}\psi^-_{\kk,\o}\equiv {1\over L^2}\sum_{\pp\not=0} 
\hat J(\pp)\d\r_{\pp,\o}\cr
&T_{- }(\hat J,\psi)={1\over L^4}
\sum_{\kk,\pp}\hat J(\pp)D_{-\o}(\pp)
\psi^+_{\kk-\pp,-\o}\psi^-_{\kk,-\o}\cr}\Eq(lillo)
$$
A crucial role in the analysis is played by the function
$$\D^{(i,j)}(\kk^+,\kk^-)
={C_{N,\o}(\kk^+,\kk^-)\over D_{-\o}(\pp)} 
g^{(i)}_\o(\kk^+) g^{(j)}_\o(\kk^-)\;.\Eq(mjmj)$$
which is such that
$$\D^{(i,j)}_\o(\kk^+,\kk^-)=0 \quad \quad i,j<N
\Eq(3.10)$$
hence at least one between $i$ or $j$ must be equal to $N$.
It holds that, for $i\le N$ 
$$\D^{N,i}(\kk^+,\kk^-)=-{f_j(\kk^-) u_N(\kk^+)\over D_\o(\kk^+-\kk^-) D_\o(\kk_-)}$$
where $u_N(\kk)=0$ for $|\kk|\le\g^N$ and $u_N(\kk)=1-f_N(\kk)$ for $|\kk|\ge \g^N$. 
It is easy to verify that
$$\D^{N,i}(\kk^+,\kk^-)={\pp\over D_{-\o}(\pp)} S^{N,i}(\kk^+,\kk^-)\Eq(lalalds)$$
with
$$|\partial^{m}_{\kk_+}\partial^{l}_{\kk_-}S^{N,i}(\kk^+,\kk^-)|
\le C_{m+l}\g^{-i(1+l)}\g^{-l(1+N)}\Eq(lalalds1)$$
from which
$$|S^{N,i}(\zz-\xx,\zz-\yy)|\le C_{n+m} {\g^N\over 1+[\g^N|\zz-\xx|]^n}
{\g^i\over 1+[\g^i|\zz-\yy|]^n}\Eq(lalalds2)$$
\vskip.5cm
\sub(1.34aklxx){\it Multiscale analysis}
The integration of $\WW_\D(J,\hat J,\phi)$ is done
by a multiscale integration similar to the previous one.
After the integration of $\psi^{(N)}$
the terms linear in $\hat J$
and quadratic in $\psi$ in the exponent
will be denoted by $K_{J}^{(N-1)}(\psi^{(\le N-1)})$;
we write $K_{J}^{(N-1)}=K_{J}^{(a,N-1)}+K_{J}^{(b,N-1)}$
where $K_{J}^{(a,N-1)}$ is obtained by the integration of 
$T_0$ and $K_{J}^{(b,N-1)}$ from the integration
of $T_-$. We can write $K_{J}^{(a,N-1)}$ as
$$K_{J}^{(a,N-1)}(\psi^{(\le N-1)})=
\sum_{\tilde \o}\int d\yy d\zz [F^{(N-1)}_{2,\o,\tilde\o}(\xx,\yy,\zz) +
F^{(N-1)}_{1,\o}(\xx,\yy,\zz) \d_{\o,\tilde\o}]
\psi^{+,\le N-1}_{\yy,\tilde\o}
\psi^{-,\le N-1}_{\zz,\tilde\o}\Eq(3.28)$$
where there is no $T_0(\hat J,\psi^{\le N-1})$ by \equ(3.10),
$F^{(N-1)}_{2,\o,\tilde\o}$ and $F^{(N-1)}_{1,\o}$ 
represent the terms in which both
or only one of the fields in $\d \r_{\pp,\o}$, respectively, are
contracted; we define
$$\LL K_{J}^{(a,N-1)}=\sum_{\tilde \o}\int d\yy d\zz 
F^{(N-1)}_{2,\o,\tilde\o}(\xx,\yy,\zz)
\psi^{+,\le N-1}_{\yy,\tilde\o}
\psi^{-,\le N-1}_{\zz,\tilde\o}\Eq(fff)$$
In the same way we decompose $K_{J}^{(b,N-1)}$ and we define $\LL$
is a similar way; the above procedure leads to two new running 
coupling functions
$$\LL K_J^{N-1}(\psi^{[\le N-1]})={1\over L^4}\sum_{\kk,\pp}
\n_{+,N-1}(\kk,\pp)\hat J(\pp)D_{\o}(\pp)
\psi^{+}_{\kk,\o} \psi^{-}_{\kk+\pp,\o} 
+\n_{-,N-1}(\kk,\pp)\hat J(\pp)D_{-\o}(\pp)
\psi^{+}_{\kk,-\o} \psi^{-}_{\kk+\pp,-\o} 
\Eq(vvn)$$
The above integration procedure can be iterated with no important
differences up to scale $h_M$; note that
$F_{1,\o}^{(k)}(\kk^+,\kk^-)$ is vanishing for $k\le N-1$.
If $W^{(h)}_{2l,m,\hat m}$ is the kernels 
in the effective potential moltiplying a monomial
in $2l$ $\psi$-fields, $m$ $J$ fields and $\hat m$
$\hat J$ fields, the following result is proved in Appendix 2.
\*
{\bf Lemma 3}{\it Assume that for $j\ge k$
$||\l_j||, ||n_j||, ||Z^{(2)}_j||$
are small enough and
$$||\n_j||\le C\l\g^{{1\over 2}(k-N)}\Eq(kkkk11)$$
then it holds the following bound,if $\hat m=0,1$
$$||W^{(h)}_{2l,m,\hat m}||\le C\l 
\g^{-h(l+m+\hat m-2)}\Eq(2.36c)$$
Moreover, if $\kk,\pp$ are fixed to an $N$ indipendent value
$$\lim_{N\to\io} R^{2,1,N}_{\o,\o',i}(\kk,\pp)=0\Eq(400.6)$$
}
\*
\vskip.5cm
\sub(1.444aklxa){\it Proof of \equ(kkkk)}

We have now to prove 
\equ(kkkk11). We can write
$$\n_{k,-}(\kk,\pp)=\n_{k,-}^a(\kk,\pp)+
\n_{k,-}^b(\kk,\pp)\Eq(eiae)$$
where in $\n^b$ are the terms obtained contracting $T_0$ and in
$\n^a$ the terms obtained contracting $T_-$.
It holds that 
$$\n_{k,-}^a(\pp,\kk)=-v(\pp)\bar\n(\pp)-v(\pp)\bar\n(\pp)<\tilde\psi(\pp);\psi^+_{\kk,-\o}
\psi^-_{\kk+\pp,-\o}>\Eq(ck6969)$$
\*
\insertplot{300pt}{100pt}%
{\ins{100pt}{55pt}{$+$}

}%
{verticiT215}{\eqg(1v)}

\* \vbox{{\ }

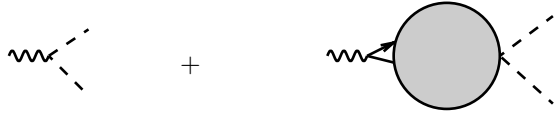
\centerline{Fig 6: Graphical representation of \equ(ck6969)}
}
\*

On the other hand 
$$\n_k^b(\pp,\kk)=
\sum_{n=0}^\io {1\over n!}
{\partial\over\partial \hat J(\pp)}
\EE^T_{k,N}(T_0;
{\partial^2\over\partial\psi^+_\kk\psi^-_{\kk+\pp}}[V...V])\Eq(fvfv67)$$
By construction the two external fields cannot be attached to $C$ (otherwise $\LL=0$). 
We distinguish now the case, as in \S 2.5, in which both the fermionic fields in $T_0$ are 
contracted with the same point with the case in which are contracted in different points. 
\*
\insertplot{300pt}{100pt}%
{\ins{68pt}{55pt}{$+$}
\ins{125pt}{77pt}{$>$}
\ins{128pt}{40pt}{$<$}
\ins{250pt}{60pt}{$+$}
}%
{verticiT216}{\eqg(1v)}

\* \vbox{{\ }

\centerline{Fig 7: Graphical representation of \equ(fvfv67); the black
dot represents  $C_N$. }
}
\*

In momentum space the first case can be written 
$$[\int {d\kk\over (2\pi)^2} {C_{\o,N}(\kk,\kk-\pp)\over D_{-\o}(\pp)}
g^{[k, N]}_\o(\kk)g^{[k, N]}_\o(\kk-\pp)] \l v(\pp) [1+<\tilde\psi(\pp) \psi^+_{\kk'}
\psi^-_{\kk'+\pp}>]\Eq(fvfv68)$$
so that summing \equ(fvfv68) with \equ(ck6969)
and using \equ(fhnk) we get a vanishing contribution
for $k\le N-|\pp|$; in fact it holds that  
$$\int {d\kk\over (2\pi)^2} {C_{\o,N}(\kk,\kk-\pp)\over D_{-\o}(\pp)}
g^{[k, N]}_\o(\kk)g^{[k, N]}_\o(\kk-\pp)] 
=\int {d\kk\over (2\pi)^2} {C_{\o,N}(\kk,\kk-\pp)\over D_{-\o}(\pp)}
g^{\le N}_\o(\kk)g^{\le N}_\o(\kk-\pp)]\Eq(fvfv69)$$
In fact $\D_{N,j}(\kk^+,\kk^-)$ is such that $|\kk^+|\ge \g^N$ 
and $|\kk^-|\ge \g^j$, then as $\pp=\kk^+-\kk^-$ necessarily
$j\ge N-|\pp|$. On the other hand for 
$k\ge N-|\pp|$
we have to prove that $||\n_k||\le \g^{-N+k}\le \g^{-|\pp|}$ which is surely true as $\pp$
is fixed.

It remain to consider the contribution to $\n_b$ 
in which the two fermionic fields in $T_0$ are contracted with different 
points. We pass to coordinate space and we proceed exactly as
in \S 2.5. 
\*
\insertplot{300pt}{100pt}%
{\ins{130pt}{60pt}{$+$}
\ins{270pt}{60pt}{$+$}
}%
{verticiT15a}{\eqg(1v)}

\* \vbox{{\ }

\centerline{Fig 8: Graphical representation of \equ(111b11)}
}
\*
Such contribution can be written as 
$$H(\xx,\yy_1,\yy_2)=\int d\zz d\zz' 
d\zz'' \sum_{i=k}^N [S^{N,i}(\xx-\zz, \xx-\zz')+S^{i,N}(\xx-\zz, \xx-\zz')]
v(\zz'-\zz''){\partial\over\partial\psi^-_{\yy_1}}{\partial\over\partial\psi^+_{\yy_2}}
{\partial^*\over\partial\psi^-_\zz}
<\psi^-_{\zz'}
\tilde\psi(\zz'')>\Eq(pov100)$$
which can be rewritten as
$$\eqalign{
&\int  d\zz d\zz' d\zz'' d\zz'''  
v(\zz'-\zz'')  \sum_{i=k}^N[S^{N,i}(\xx-\zz, \xx-\zz')+S^{i,N}(\xx-\zz, \xx-\zz')]
{\partial\over\partial\psi^-_{\yy_1}}{\partial\over\partial\psi^+_{\yy_2}}
{\partial^*\over\partial\psi_\zz}
<\psi^-_{\zz'};
\tilde\psi(\zz'')>
+\cr
&\int d\zz d\zz' d\zz''
\sum_{i=k}^N  [S^{N,i}(\xx-\zz, \xx-\zz')+S^{i,N}(\xx-\zz, \xx-\zz')]
H_{2,0}(\zz',\zz'')g^{(k,N)}_{\o,\o}(\zz''-\zz)]
v(\zz-\zz''')H_{1,2}(\zz''',\yy_1,\yy_2)\cr}
\Eq(111b11)$$
We will prove in the Appendix that the first addend in \equ(111b11)
is bounded by
$$||\sum_{i=k}^N [S^{N,i}(\xx-\zz, \xx-\zz')+S^{i,N}(\xx-\zz, \xx-\zz')]
{\partial^*\over\partial\psi_\zz}
<\psi^-_{\zz'};
\tilde\psi(\zz'');\tilde\psi(\yy)>||\le C\l\g^{-{1\over 2}(N+k)}
\Eq(jjjn)$$
On the other hand the second addend in \equ(111b11) is easily bounded by noting that, by momentum
conservation,
$$
||\sum_{i=k}^N  [S^{N,i}(\xx-\zz, \xx-\zz')g^{(i)}_{\o,\o}(\zz''-\zz)
+S^{i,N}(\xx-\zz, \xx-\zz')g^{(N)}_{\o,\o}(\zz''-\zz)]
H_{2,0}(\zz',\zz'')v(\zz-\zz''')H_{1,2}(\zz''',\yy_1,\yy_2)||\Eq(laii)$$
%
We proceed as in \S 2.5, and in the first addend we integrate
over the line $\xx-\zz$, using that $||H_{2,0}||\le C\l\g^{-k}$ and 
$|g^{(i)}_{\o,\o}|\le C \g^{-i}$
getting $\sum_{i=k}^N \g^{-N}\g^{-i}\g^i\g^{-k}\le |N-k|\g^{-N-k}\le C \g^{(-N-k)/2}$;
in the second addend we integrate
over the line $\xx-\zz'$, getting $\sum_{i=k}^N \g^{-2 N}\g^{i}\g^{-k}\le C \g^{(-N-k)/2}$.

A similar analysis can be done for $\n_{k,+}$ with the only difference that the first
term in \equ(ck6969) and in Fig.7 is absent.
\*
\section(2a,Appendix 1)
\vskip.5cm
\vskip.5cm
\sub(77){\it Proof of Lemma 1}
For an introduction to the formalism used in this section
we will refer to [GM]. We define a family of trees in the following way.

\*
\insertplot{300pt}{200pt}%
{{\ins{30pt}{85pt}{$r$}\ins{50pt}{85pt}{$v_0$}\ins{130pt}{100pt}{$v$}%

\ins{35pt}{-2pt}{$h$}\ins{55pt}{-2pt}{$h+1$}\ins{135pt}{-2pt}{$h_v$}%

\ins{235pt}{-2pt}{$N$}\ins{255pt}{-2pt}{$N+1$}}%

}%
{treelut2}{\eqg(1vq)}
\* \vbox{{\ }
\*
\*
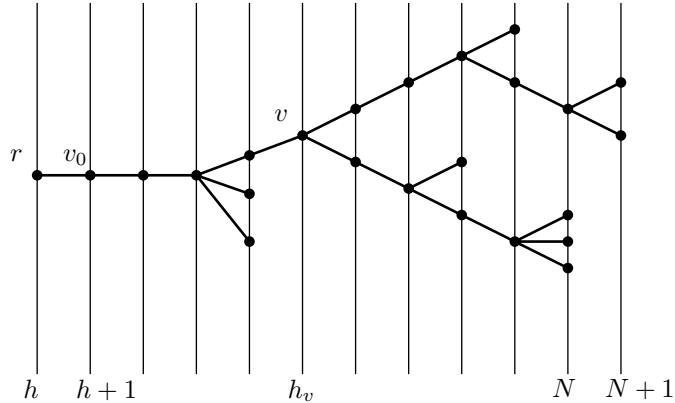
\centerline{Fig 9: an example of tree $\t$}
}
\*
\*
\0 1) Let us consider the family of all trees which can be
constructed by joining a point $r$, the {\it root}, with an
ordered set of $n$ points, the {\it endpoints} of the {\it
unlabelled tree}, so that $r$ is not a branching
point. $n$ will be called the {\it order} of the unlabelled tree
and the branching points will be called the {\it non trivial
vertices}. The unlabelled trees are partially ordered from the
root to the endpoints in the natural way; we shall use the symbol
$<$ to denote the partial order.

Two unlabelled trees are identified if they can be superposed by a
suitable continuous deformation, so that the endpoints with the
same index coincide. It is then easy to see that the number of
unlabelled trees with $n$ end-points is bounded by $4^n$.

We shall consider also the {\it labelled trees} (to be called
simply trees in the following); they are defined by associating
some labels with the unlabelled trees, as explained in the
following items.

\0 2) We associate a label $h\le N-1$ with the root and we denote
by $\TT_{h,n}$ the corresponding set of labelled trees with $n$
endpoints. Moreover, we introduce a family of vertical lines,
labelled by an integer taking values in $[h,N+1]$, and we represent
any tree $\t\in\TT_{h,n}$ so that, if $v$ is an endpoint or a non
trivial vertex, it is contained in a vertical line with index
$h_v>h$, to be called the {\it scale} of $v$, while the root is on
the line with index $h$. There is the constraint that, if $v$ is
an endpoint, $h_v>h+1$.

The tree will intersect in general the vertical lines in set of
points different from the root, the endpoints and the non trivial
vertices; these points will be called {\it trivial vertices}. The
set of the {\it vertices} of $\t$ will be the union of the
endpoints, the trivial vertices and the non trivial vertices. Note
that, if $v_1$ and $v_2$ are two vertices and $v_1<v_2$, then
$h_{v_1}<h_{v_2}$.

Moreover, there is only one vertex immediately following the root,
which will be denoted $v_0$ and can not be an endpoint (see
above); its scale is $h+1$.

Finally, if there is only one endpoint, its scale must be equal to
$h+2$.

\0 3) With each endpoint $v$ of scale $h_v=N+1$ we associate one of the
monomials in the exponential of \equ(gf)  and a set
$\xx_v$ of space-time points (the corresponding integration
variables); with each
endpoint of scale $h_v\le N$ we associate one of contributions 
in $\LL\VV^{(h_v)}$ \equ(2.22).
We impose the constraint that, if $v$ is an endpoint,
$h_v=h_{v'}+1$, if $v'$ is the non trivial vertex immediately
preceding $v$. Given a vertex $v$, which is not an endpoint, $\xx_v$ will denote
the family of all space-time points associated with one of the
endpoints following $v$.

\0 4) The trees containing only the root and an endpoint of scale
$h+1$ (note that they do not belong to $\TT_{h,N+1}$ ) will be
called the {\it trivial trees}.

In terms of these trees, the effective potential $\VV^{(h)}$,
$h\le 1$, can be written as
$$\VV^{(h)}(\psi^{(\le h)}) + L^2 \tilde F_{h+1}=
\sum_{n=1}^\io\sum_{\t\in\TT_{h,n}} \VV^{(h)}(\t,\psi^{(\le
h)})\;,\Eq(2.61)$$
where, if $v_0$ is the first vertex of $\t$ and $\t_1,..,\t_s$
($s=s_{v_0}$) are the subtrees of $\t$ with root $v_0$,
$\VV^{(h)}(\t,\psi^{(\le h)})$ is defined inductively by the
relation
$$ \VV^{(h)}(\t,\psi^{(\le h)})= {(-1)^{s+1}\over s!}
\EE^T_{h+1}[ \bar\VV^{(h+1)}(\t_1,\psi^{(\le h+1)});\ldots;
\bar\VV^{(h+1)}(\t_{s},\psi^{(\le h+1)})]\;,\Eq(2.62)$$
and $\bar\VV^{(h+1)}(\t_i,\psi^{(\le h+1)})$,if $\RR=1-\LL$

\0 a) is equal to $\RR\VV^{(h+1)}(\t_i,\psi^{(\le h+1)})$ if the
subtree $\t_i$ is not trivial;

\0 b) if $\t_i$ is trivial and $h< N-1$, it is equal to
$\LL\VV^{(h+1)}$ or, if $h=N-1$, to one of the monomials
contributing to $\VV^{(N)}(\psi^{\le N})$.

\0 $\EE^T_{h+1}$ denotes the truncated expectation with respect to
the measure $P(d \psi^{(h+1)})$
We associate then to each vertex $v$ is associated the set $P_v$, the set of
labels of {\it external fields of $v$}, that is
the field variables of type $\psi$ which belong to one of the
endpoints following $v$ and either are not yet contracted in the
vertex $v$ (we call $P_v^{(n)}$ the set of these variables)
or are contracted with the $\psi$ variable of an endpoint of type
$\f$ through a propagator $g^{[h_v,N]}$.
The sets $|P_v|$ must
satisfy various constraints. First of all, if $v$ is not an
endpoint and $v_1,\ldots,v_{s_v}$ are the vertices immediately
following it, then $P_v \subset \cup_i P_{v_i}$; if $v$ is an
endpoint, $P_v$ is the set of field labels associated to it.
We shall denote $Q_{v_i}$ the intersection of
$P_v$ and $P_{v_i}$; this definition implies that $P_v=\cup_i
Q_{v_i}$. The subsets $P_{v_i}\bs Q_{v_i}$, whose union ${\cal
I}_v$ will be made, by definition, of the {\it internal fields} of
$v$, have to be non empty, if $s_v>1$. Given $\t\in\TT_{h,n}$,
there are many possible choices of the
subsets $P_v$, $v\in\t$, compatible with all the constraints. We
shall denote $\PP_\t$ the family of all these choices and $\bP$
the elements of $\PP_\t$.
Moreover, we associate with any $f\in {\cal I}_v$ a scale label
$h(f)=h_v$. We call $\chi$-vertices the vertices of $\t$ such that their set 
${\cal I}_v$ of internal lines is not empty; $V_\chi(\t)$ will denote the set of all
$\chi$-vertices of $\t$.

With these definitions, we can rewrite $\VV^{(h)}(\t,\psi^{(\le h)})$
in the r.h.s. of \equ(2.61) as:
$$\eqalign{&\VV^{(h)}(\t,\psi^{(\le
h)})=\sum_{\bP\in\PP_\t}
\VV^{(h)}(\t,\bP)\;,\cr &\VV^{(h)}(\t,\bP)=\int d\xx_{v_0}
\tilde\psi^{(\le h)}(P_{v_0})
K_{\t,\bP}^{(h+1)}(\xx_{v_0})\;,\cr}\Eq(2.64)$$
where
$$\tilde\psi^{(\le h)}
(P_{v})=\prod_{f\in
P_v}\psi^{ (\le
h)\e(f)}_{\xx(f),\o(f)}\Eq(2.65)$$
and $K_{\t,\bP}^{(h+1)}(\xx_{v_0})$ is defined inductively by
the equation, valid for any $v\in\t$ which is not an endpoint,
$$K_{\t,\bP,\O}^{(h_v)}(\xx_v)={1\over s_v !}
\prod_{i=1}^{s_v} [K^{(h_v+1)}_{v_i}(\xx_{v_i})]\; \;\EE^T_{h_v}[
\tilde\psi^{(h_v)}(P_{v_1}\bs Q_{v_1}),\ldots,
\tilde\psi^{(h_v)}(P_{v_{s_v}}\bs
Q_{v_{s_v}})]\;,\Eq(2.65)$$
where 
$\tilde\psi^{(h_v)}(P_{v_i}\bs Q_{v_i})$ has a definition
similar to \equ(2.65). Moreover, if $v$ is an endpoint and $h_v\le
N$, $K^{(h_v)}_v(\xx_v)= \l_{h_v},n_{h_v},1+Z^{(2)}_{h_v}$, while if $h_v=N+1$
$K^{(1)}_v$ is equal to one of the kernels of the monomials in
(2.1).

\equ(2.61)--\equ(2.64) is not the final form of our expansion;
we further decompose $\VV^{(h)}(\t,\bP)$, by using the
following representation of the truncated expectation in the
r.h.s. of
\equ(2.65). Let us put $s=s_v$, $P_i\=P_{v_i}\bs Q_{v_i}$;
moreover we order in an arbitrary way the sets $P_i^\pm\=\{f\in
P_i,\e(f)=\pm\}$, we call $f_{ij}^\pm$ their elements and we
define $\xx^{(i)}=\cup_{f\in P_i^-}\xx(f)$, $\yy^{(i)}=\cup_{f\in
P_i^+}\xx(f)$, $\xx_{ij}=\xx(f^-_{i,j})$,
$\yy_{ij}=\xx(f^+_{i,j})$. Note that $\sum_{i=1}^s
|P_i^-|=\sum_{i=1}^s |P_i^+|\=n$, otherwise the truncated
expectation vanishes. A couple
$l\=(f^-_{ij},f^+_{i'j'})\=(f^-_l,f^+_l)$ will be called a line
joining the fields with labels $f^-_{ij},f^+_{i'j'}$
connecting the points
$\xx_l\=\xx_{i,j}$ and $\yy_l\=\yy_{i'j'}$, the {\it endpoints} of
$l$. Then, it is well known [Le, GM] that, up to
a sign, if $s>1$,
$$\eqalign{
&\qquad \EE^T_{h}(\tilde\psi^{(h)}(P_1)\tilde\psi^{(h)}(P_2)
...\tilde\psi^{(h)}(P_s))=\cr
&= \sum_{T}\prod_{l\in T} \tilde g^{(h)}(\xx_l-\yy_l)
\int dP_{T}(\tt) \det
G^{h,T}(\tt)\;,\cr}\Eq(2.66)$$
where $T$ is a set of lines forming an {\it anchored tree graph} between
the clusters of points $\xx^{(i)}\cup\yy^{(i)}$, that is $T$ is a
set of lines, which becomes a tree graph if one identifies all the
points in the same cluster. Moreover $\tt=\{t_{i,i'}\in [0,1],
1\le i,i' \le s\}$, $dP_{T}(\tt)$ is a probability measure with
support on a set of $\tt$ such that $t_{i,i'}=\uu_i\cdot\uu_{i'}$
for some family of vectors $\uu_i\in \RRR^s$ of unit norm. Finally
$G^{h,T}(\tt)$ is a $(n-s+1)\times (n-s+1)$ matrix, whose elements
are given by
$$G^{h,T}_{ij,i'j'}=t_{i,i'} \tilde
g^{(h)}(\xx_{ij}-\yy_{i'j'})\Eq(2.67a)$$
with $(f^-_{ij}, f^+_{i'j'})$ not belonging to $T$.

In the following we shall use \equ(2.66) even for $s=1$, when $T$
is empty, by interpreting the r.h.s. as equal to $1$, if
$|P_1|=0$, otherwise as equal to $\det
G^{h}=\EE^T_{h}(\tilde\psi^{(h)}(P_1))$. \*

If we apply the expansion \equ(2.66) in each non trivial vertex of
$\t$, we get an expression of the form
$$ \VV^{(h)}(\t,\bP) = \sum_{T\in {\bf T}} \int d\xx_{v_0}
\tilde\psi^{(\le h)}(P_{v_0}) W_{\t,\bP
,T}^{(h)}(\xx_{v_0}) \= \sum_{T\in {\bf T}}
\VV^{(h)}(\t,\bP,T)\;,\Eq(2.68)$$
where ${\bf T}=\bigcup_v T_v$. Given $\t\in\TT_{h,n}$ and the labels $\bP,T$,
calling $v_i^*,\ldots,v_n^*$ the endpoints of $\t$ and putting
$h_i=h_{v_i^*}$, we get the bound
$$\eqalign{&
|W_{\t,\bP, T}(\xx_{v_0})|\le \int\prod_{l\in T^*}
d(\xx_l-\yy_l) \left[\prod_{i=1}^n |v_{h_i-1}
(\xx_{v_i^*})|\right]\cdot\cr
&\qquad\qquad\cdot\Big\{\prod_{v\,\atop\hbox{\rm not
e.p.}}{1\over s_v!} \max_{\tt_v}|\det
G^{h_v,T_v}(\tt_v)|\prod_{l\in T_v}|
g^{(h_v)}(\xx_l-\yy_l)|\Big\}\;,\cr}\Eq(2.79)$$
where $T^*$ is a tree graph obtained from $T=\cup_vT_v$, by adding
in a suitable (obvious) way, for each endpoint $v_i^*$,
$i=1,\ldots,n$, the lines connecting the space-time points
belonging to $\xx_{v_i^*}$.
A standard application of Gram--Hadamard inequality, combined with
the dimensional bound on $g^{(h)}$ 
implies that
$$|\det G_\a^{h_v,T_v}(\tt_v)| \le
c^{\sum_{i=1}^{s_v}|P_{v_i}|-|P_v|-2(s_v-1)}\cdot\;
\g^{{h_v}\left(\sum_{i=1}^{s_v}|P_{v_i}|-|P_v|-2(s_v-1)\right)}\;.
\Eq(2.80)$$
Moreover
$$\prod_{v\,\hbox{\ottorm not e.p.}}
{1\over s_v!}\int\prod_{l\in T^*}
d(\xx_l-\yy_l) \prod_{i=1}^n |v_{h_i-1}(\xx_{v_i^*})|
|g^{(h_v)}_{\o_l}(\xx_l-\yy_l)|\le c^n 
\prod_{v\,{\rm not e.p.}} {1\over s_v!} \g^{-h_v(s_v-1)}\;.\Eq(2.81)$$
so we can bound the r.h.s. of \equ(2.79) by,if $\bar n+n_J+n_\phi=n$
$$(c\l)^{\bar n}
\g^{h(2-{|P_{v_0}|\over 2}-n^J_{v_0}-n_{2,v_0})}
\prod_{v\ {\rm not}\ {\rm e. p.
}}{1\over s_v!}\g^{-({|P_v|\over 2}-2+n^J_v+n_{2,v})}\Eq(rrr)$$
where $n^J_v$ are the $J$ fields associated to the endpoints following $v$, $n_{2,v}$
is the number of endpoints of type $n_{k}$ following $v$ and $\bar n+n_J+n_\phi=n$.
We can bound \equ(rrr) by
$$(c\l)^{\bar n} \g^{-h d_{v_0}}
\prod_{\tilde v\in V_\c(\t)}
{1\over s_{\tilde v}!}\g^{-(h_{\tilde v}-h_{\tilde v'})d_{\tilde v}}\Eq(rrr1)$$
where $d_v$ is the {\it dimension}, 
$d_v={|P_v|\over 2}-2+n^J_v+n_{2,v}$ and
and $\tilde v'$ is the $\c$-vertex immediately preceding $\tilde v$.
By construction the vertices $v$ with $|P_v|=2,4$ and 
$n^J_v=0$, or $|P_v|=2$ and $n^J_v=1$ are necessarily endpoints so they
do not belong to $V_\c$, hence $d_v\ge 1$.
In order to sum over $\t$ and $\bP$
we note that
the number of unlabeled trees is $\le
4^n$; fixed an unlabeled tree, the number of terms in the sum over the
various labels of the tree is bounded by $C^n$, except the sums over the scale
labels and the sets $\bP$. 
Regarding the sum over $T$, it is empty if $s_v=1$. If $s_v>1$ and
$N_{v_i}\=|P_{v_i}|-|Q_{v_i}|$,
the number of anchored trees with $d_i$ lines branching from the vertex
$v_i$ can be bounded, by using Caley's formula, by
$${(s_v-2)!\over (d_1-1)!...(d_{s_v}-1)!} N_{v_1}^{d_1}...
N_{v_{s_v}}^{d_{s_v}}\;;\Eq(gfggh)$$
hence the number of addenda in $\sum_{T\in {\bf T}}$ is bounded by
$\prod_{v\,\hbox{not \ottorm e.p.}} s_v!\;
C^{\sum_{i=1}^{s_v}|P_{v_i}|-|P_v|}$.
 
In order to bound the sums over the scale labels and $\bP$ we first use
the inequality, 
following from \equ(rrr1)
$$\prod_{v\in V_\chi(\t)}\g^{-(h_v-h_{v'})
d_v}\le [\prod_{v\in V_\chi(\t)} 
\g^{-{1\over 40}(h_{v}-h_{v'})}]
[\prod_{v\in V_\chi(\t)}\g^{-{|P_v|\over 40}}]\Eq(3.111)$$
The
factors $\g^{-{1\over 40}(h_{\tilde v}-h_{\tilde v'})}$ in the r.h.s.
of \equ(3.111) allow to bound the sums over the scale labels by $C^n$.
The sum over $\bP$ can be bounded by using the following combinatorial
inequality. Let $\{p_v, v\in \t\}$ a set of integers such that
$p_v\le \sum_{i=1}^{s_v} p_{v_i}$ for all $v\in\t$ which are not endpoints;
then (see for instance App. 6 of [GM])
$$\sum_{\bP}\prod_{v\in V_\chi(\t)}
\g^{-{|P_v|\over 40}}\le \prod_{v\in V_\chi(\t)} \sum_{p_v}
\g^{-{p_v\over 40}} B(\sum_{i=1}^{s_v}p_v,p_v) \le C^n\;.\Eq(3.113)$$
where $B(n,m)$ is the binomial coefficient. This concludes the proof of the Lemma.
\*
{\it Remark 1} If in $\t$ there are two $\c$-vertices with scale $h_1$ and $h_2$,
we can write the r.h.s. of \equ(rrr1) as
$$(c|\l|)^n \g^{-h d_{v_0}}\g^{-{1\over 2}|h_1-h_2|}
\prod_{\tilde v\in V_\c(\t)}
{1\over s_{\tilde v}!}\g^{-{1\over 2}(h_{\tilde v}-h_{\tilde v'})d_{\tilde v}}\Eq(rrr1)$$
as $d_{\tilde v}\ge 1$; of course the sum over $\t,\bP$
can be performed as above and
this implies that the dimensional bound \equ(2.36)
can be improved by $\g^{-{1\over 2}|h_1-h_2|}$ for such trees; this property is called 
{\it short memory property}.
\*
{\it Remark 2} Let us consider a tree $\t$ contributing
to a kernel $W^{(h)}_{2l,m}$ for which $\LL=1$ (see \equ(2.20));
then $v_0\in V_\chi$; in fact if $v_0\not\in V_\chi$
then $v_0$ is trivial and the external
fields of $v_0$ and $v_1$ are the same,if $v_1$ is the vertex preceding $v_0$;
then $\RR\VV(\t_1,\psi)=0$. 
\vskip.5cm
\sub(1.22aklxa){\it Proof of \equ(2.36a)}
It holds that
$${\partial^*\over\partial\psi^-_\yy} <\psi^-_\xx;
\tilde\psi(\tilde\yy)>_T
=
{\partial^2\over\partial\phi(\yy)\phi(\xx)
}{\partial^2\over\partial  J(\tilde\yy) \bar J(\xx)}
\HH(\phi,J,\bar J)\Eq(gffeff)$$
where
$$e^{\HH(\phi,J,\bar J)}=
\int P_{k,N}(d\psi)e^{-\VV(\psi+\phi)+\sum_\o \int d\zz 
J(\zz)[\psi^+_{\zz,\o}+\phi^+_{\zz,\o}]
[\psi^-_{\zz,\o}+\phi^-_{\zz,\o}]+\int \bar J(\zz)\phi(\zz)
[\psi_\zz+\phi_\zz]}\Eq(gfffe1)$$
%
%
%
where the derivative over $\phi$ cannot 
be applied over $[\psi_\zz+\phi_\zz]$ otherwise a disconnected
contribution is found. \equ(gfffe1)
can be integrated by a multiscale analysis as \equ(1);
after the integration of the scales $N,N-1,..,h$ we get
$\int P_{k,h}(d\psi)e^{-\bar\VV^{(h)}(\psi+\phi)}$ with
$$\bar\VV^{(h)}(\psi)=\sum_{l,\underline\o,\underline\e}\int 
d\xx_1...d\xx_{2l} W^{(h)}_{2l,m,\bar m,\underline\o,\underline\e}
\prod_{i=1}^{2 l} \psi^{\e_i}_{\xx_i,\o_i}
\prod_{i=1}^m J_{\o_i}(\xx_i)\prod_{i=1}^{\bar m}\bar J_{\o_i}(\xx_i)\Eq(2.17a)
$$
and 
$${\partial^*\over\partial\psi^-_\yy} <\psi^-_\xx;
\tilde\psi(\tilde\yy)>_T
=\sum_{h=k}^N W^{(h)}_{2,1,1}\Eq(pappa1)$$
We integrate $\WW(\phi,J,\bar J)$
by a multiscale procedure identical to the one for 
$\WW(\phi,J)$ (in particular $\LL W^{(h)}_{2l,m,\bar m}$
if $\bar m\not=0$, so that no new running coupling functions are
introduced) 
and we still get the bound 
\equ(rrr1) in which $d_v$ is given by 
${|P_v|\over 2}-2+n^J_v+n^{\bar J}_v+n_{2,v}$.
There is an apparent problem due to the fact that 
$d_v=0$ for the vertices
with one external 
$\bar J$ line and with two external fermionic lines.
Let us consider the terms
$\int d\kk d\pp W^{(h_v)}_{2,0,1}(\kk,\pp)\bar J(\pp)\phi^+_\kk\psi^-_{\kk+\pp}$
associated with such vertices;
of course $W^{(h_v)}_{2,0,1}=g^{[h_v,N]}(\kk+\pp) G_2(\kk+\pp)$; the momentum $\kk+\pp$
of the external $\psi$ fields has scale $\g^{h_{v'}}$, and 
$g^{h_{v'}}(\kk+\pp)g^{h_{v}}(\kk+\pp)$
is nonvanishing only if $|h_{v'}-h_{v}|\le 2$;
hence, as
$\g^{h_v-h_{v'}}\le \g^2$, we can replace $d_v$ with
$d_v+\e_v$,with $\e_v=1$ when $|P_v|=2,n^{\bar J}_v=1$ so that we get
$||W^{(h)}_{2l,m,\bar m}||\le C\l
\g^{-h(l+m+\bar m-2)}$ implying $||W^{(h)}_{2,1,1}||\le C\l\g^{-h}$.
\*
\insertplot{300pt}{100pt}%
{
}%
{verticiT111}{\eqg(1v)}

\* \vbox{{\ }

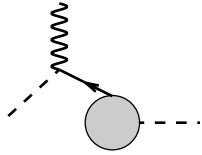
\centerline{Fig 10: Graphical representation of marginal terms $\bar J \phi \psi$}
}
\*
In the same way
$${\partial^*\over\partial\psi^-_\yy}
<\psi^-_\xx;\tilde\psi(\tilde\yy);\tilde\psi(\zz)>_T=
{\partial^2\over\partial\phi(\yy)\phi(\xx)
}{\partial^2\over\partial  J(\tilde\yy) J(\zz)\bar J(\xx)}
\HH(\phi,J,\bar J)\Eq(gffeff)$$
and $||W^{(h)}_{2,2,1}||\le C\l\g^{-2h}$ from which \equ(lau1) follows.
Finally \equ(2.35g) follows fromthe fact that 
$||W^{(h)}_{3,2,1}||\le C\l\g^{-2h}$.

\vskip.5cm
\sub(1.23aklxa){\it Proof of \equ(2.36b)}

We can write
%
$$\int d\zz g^{[k,N]}(\xx-\zz') g^{[k,N]}(\xx-\zz) 
{\partial^*\over\partial\psi_\zz}
<\psi^-_{\zz'};
\tilde\psi(\zz'');\tilde\psi(\yy)>=$$
$$={\partial\over\partial \bar J(\zz') \bar J(\xx)}{\partial\over\partial  J(\zz'')
\partial J(\yy)} \log 
\int P^{[k,N]}_\s(d\psi)P^{[k,N]}_{-\s}(d\psi)
e^{-\VV(\psi_\s)}
e^{\int d\zz[\sum_\e \bar J(\zz)\psi^\e_{\s,\zz}\psi^{-\e}_{\s,\zz}+\bar J(\zz) \psi^{\e}_{\s,\zz}
\psi^{-\e}_{-\s,\zz}]}\Eq(vvv)$$
where $\VV(\psi)=-\l {1\over 2}
 \sum_{\o=\pm}\int\!d\xx\ 
 \hp^{(\le N)+}_{\xx,\o}
\hp^{(\le N)-}_{\xx,\o}\hp^{(\le N)+}_{\xx,-\o}\hp^{(\le N)-}_{\xx,-\o}$
and $\psi_\s,\psi_{-\s}$ are indipendent
fields. In order to check \equ(vvv) we note that the r.h.s. of \equ(vvv)
gives
$$<\psi^+_{\s,\zz'}\psi^-_{-\s,\zz'};\psi^-_{\s,\xx}\psi^+_{-\s,\xx};
\tilde\psi(\zz'');\tilde\psi(\yy)>=g^{[k,N]}(\xx-\zz')
<\psi^+_{\zz'};\psi^-_{\xx};
\tilde\psi(\zz'');\tilde\psi(\yy)>
\Eq(200.3)$$
and we use the identity $\psi^-_{\xx}=\int d\zz g^{[k,N]}(\xx-\zz)
{\partial^*\over\partial\psi_\zz}$, where we have used that 
${\partial\over\partial\psi_\zz}$ cannot be applied over 
$\psi^+_{\zz'}$ otherwise a disconnected contribution is found.
The r.h.s.of \equ(vvv)
can be integrated by a multiscale analysys as above,
and after the integration of the scales $N,N-1,..k$
and the effective potential have the form
\equ(2.17a) and kernels $W^k_{l_1,l_2,m_1,m_2}$
multiplying $l_1$ fields $\psi_\s$, $l_1$ fields $\psi_{-\s}$,
$m_1$ fields $J$ and $m_2$ $\bar J$. There are new terms
with vanishing dimension: there are no 
vertices with external lines $\bar J \psi^+_\s\psi^-_\s$,
as the contraction of the fields $\psi_{-\s}$ means that
there are at least two external lines $\bar J$;
the vertices with external lines $\bar J \psi^+_\s\psi^-_{-\s}$
have surely a propagator with the same momentum as the external line,
then $\g^{h_v-h_v'}\le\g^2$ and the sum over trees can be done
without introducing anynew coupling. Then the norm of the l.h.s. of \equ(vvv)
is bounded by $\sum_{h=k}^N ||W^k_{0,0,2,2}||
\le C\l\g^{-2h}$ so that we get \equ(2.36b)
\vskip1cm
\section(3a,Appendix 2)
\*
\vskip.5cm
\sub(1.24aklxa){\it Proof of Lemma 3}
As the case $\hat m=0$ is identical to the previous one, we consider
only the case $\hat m=1$. The trees are essentially
identical with the ones in Appendix 1, with the only difference
that there is or an end-point associated to $\n_{j,\pm}$ at scale $j$,
or an endpoint associated to $T_0$.
In the first case $d_v\ge 1$ for any $v$ by construction. In the second case
there is surely
a $\c$-vertex at scale $N-1$, by \equ(3.10);
moreover the only vertex $v$ with $d_v=0$ has
one external line $\hat J$ and two external $\psi$-lines;
it  has necessarily scale $N-1$ and the form
$$F_{1,\o}^{(N-1)}(\kk^+,\kk^-)=
[{[C_{N}(\kk^-)-1] D_\o(\kk^-) \hat g^{(N)}_+(\kk^+)- u_N(\kk^+)
\over D_\o(\kk^+ -\kk^-)} G^{(2)}(\kk^+)\Eq(3.28a)$$
for a suitable function $G^{(2)}(\kk)$; by the support properties 
of the functions $\hat g^{(N)}_+(\kk^+)$, $u_N(\kk^+)$,
there is a nonvanishing contribution only if
the external line with momentum $\kk^+$ is contracted at scale $N-2$,
so that $\g^{h_{v}-h_{v'}}\le\g^2$. Hence we get 
$$||\WW_{\t}||\le C^n \l^n
\g^{-h(2-{n_\phi\over 2}-n_{\hat J}-n_J)}
\prod_{v}\g^{-{1\over 2}(h_v-h_{v'})(-2+{|P_v|\over 2}+n^{\hat J}_v
+n^J_v+\e_v)}\Eq(ddcc)$$
where $\e_v=1$ if $|P_v|=2, n^{\hat J}_v=1$. The sum
over $\t$ can be done as in Appendix 1 and \equ(2.36c) is found.
Moreover \equ(400.6) follows noting that
the trees contributing to 
$R^{2,1,N}_{\o,\o',i}(\kk,\pp)$ have an endpoint associated to $T_0$
or an $\n_{j,\pm}$ at scale $j$.
There is then a gain,
with respect to the dimensional bound, of a factor
$\g^{{1\over 2}(\bar h-N)}$, if $\bar h$ is the scale of $\kk$;
in fact the trees contributing to $R^{2,1,N}_{\o,\o',i}(\kk,\pp)$
have surely a $\c$-vertex at scale
$\bar h$, and an end-point associated to $\n_{j,\pm}$ at scale $j$,
or an endpoint associated to $T_0$; by the short memory property and \equ(kkkk11)
it follows \equ(400.6).
\*
\sub(1.24aklxa){\it Proof of \equ(jjjn)} We can write
$$\eqalign{
&H^{(k)}_{0,0,2,2}(\zz',\xx,\zz'',\yy_1,\yy_2)=
[S^{N,i}(\xx-\zz, \xx-\zz')+S^{i,N}(\xx-\zz, \xx-\zz')]
{\partial\over\partial\psi^-_{\yy_1}}{\partial\over\partial\psi^+_{\yy_2}}
{\partial^*\over\partial\psi_\zz}
<\psi^-_{\zz'};
\tilde\psi(\zz'')>=\cr
&{\partial\over\partial \hat J(\zz') \bar J(\xx)}{\partial\over\partial  J(\zz'')} 
{\partial^2\over\partial\phi(\yy_1)\partial\phi(\yy_2)}
\log 
\int P^{[k,N]}_\s(d\psi)P^{[k,N]}_{-\s}(d\psi)
e^{-\VV(\psi_\s)}
e^{\int\sum_\e \hat J(\zz)\d\r(\zz)+\bar J(\zz) \psi^{\e}_{\s,\zz}
\psi^{-\e}_{-\s,\zz}}
\cr}\Eq(vvvoo)
$$
where $\VV(\psi)=-\l {1\over 2}
 \sum_{\o=\pm}\int\!d\xx\ 
 \hp^{(\le N)+}_{\xx,\o}
\hp^{(\le N)-}_{\xx,\o}\hp^{(\le N)+}_{\xx,-\o}\hp^{(\le N)-}_{\xx,-\o}$
and 
$$\d\r(\zz)=\int d\pp e^{i\pp\zz}\int d\kk C_N(\kk,\kk+\pp)
\psi^+_{\kk,\s}\psi^-_{\kk+\pp,-\s}$$
The r.h.s.of \equ(vvvoo)
can be integrated by a multiscale analysys as above,
and after the integration of the scales $N,N-1,..k$
and the effective potential have the form
\equ(2.17a) and kernels $W^k_{l_1,l_2,m_1,m_2}$
multiplying $l_1$ fields $\psi_\s$, $l_1$ fields $\psi_{-\s}$,
$m_1$ fields $J$ and $m_2$ $\bar J$. There are new terms
with vanishing dimension: there are no 
vertices with external lines $\bar J \psi^+_\s\psi^-_\s$,
as the contraction of the fields $\psi_{-\s}$ means that
there are at least two external lines $\bar J$;
the vertices with external lines $\bar J \psi^+_\s\psi^-_{-\s}$
have surely a propagator with the same momentum as the external line,
then $\g^{h_v-h_{v'}}\le\g^2$ and the sum over trees can be done
without introducing any new coupling. 

We can write (analogously to \equ(marmat))
$$H^{(k)}_{2,0,1,2}=\sum_{i=k}^N
\sum_{\t\in\TT^*_i}W^{(k)}_{\t;2,0,1,2}\Eq(laolo)$$
where $\TT^*_i$ is the set of trees with root at scale $i$
and such that $v_0$ is a $\c$-vertex. Note that, by construction, there is surely a $\c$
vertex at scale $N$, hence the dimensional bound is improved by 
a factor $\g^{{1\over 2}(i-N)}$,see \S 5.1, so that
$$||H^{(k)}_{2,0,1,2}||=\sum_{i=k}^N
C \l^2\g^{-2i} \g^{{1\over 2}(i-N)}\le \tilde C\l^2 \g^{-k} \g^{-{1\over 2}N}\Eq(gbvb)
$$
\vskip.5cm
\sub(1.34aklxa){\it The limit of local interaction.}
In order to prove Theorem 2, we consider
$v(\pp)=e^{-\pp^2\over K^2}$ with suitable $Z_K,m_K$
and $\g^{-h_M}=K$. We fix $K$ and then we proceed as above
by taking the limit $N\to\io$, and from the previous analysis it follows that the
WI verifyes \equ(4001). We consider then the limit $K\to\io$;
the contribution of a tree with a vertex with scale $h_v\ge h_M$
to a Schwinger function with fixed coordinate is vanishing as $K\to\io$,
by the short memory property, and by \equ(rcc100) the bare parameters have to be chosen as in 
\eq(44).
\*
\section(3a,Appendix 3:Perturbative Computations)
\*
\vskip.5cm
We can check the WI \equ(90), (1.15), (1.19)
by a naive perturbative computation at lowest orders.
Note that 
$j^0_\zz=\sum_{\o'=\pm}\psi_{\o',\xx}^+\psi_{\o',\xx}^-$
and $j^1_\zz=i\sum_{\o'=\pm}\o'\psi_{\o',\xx}^+\psi_{\o',\xx}^- $
and therefore
$\pp_\m \hat j^\m_\pp=i\sum_{\o'=\pm} D_{\o'}(\pp) \hat\r_{\o',\pp}$,
$\pp_\m \hat j^{5,\m}_\pp=i\sum_{\o'=\pm}\o' D_{\o'}(\pp) \hat\r_{\o',\pp}$
so that
$$\eqalign{
D_{\o'}(\pp) \la\hat \r_{\o',\pp}\hp^-_{\o,\kk}\hp^+_{\o,\kk+\pp}\ra
=&
\d_{\o,\o'}\left[\la\hp^-_{\o,\kk}\hp^+_{\o,\kk}\ra-
\la\hp^-_{\o,\kk+\pp}\hp^+_{\o,\kk+\pp}\ra\right]\cr
&+\n_+(\pp)
D_{\o'}(\pp) \la\hat \r_{\o',\pp}\hp^-_{\o,\kk}\hp^+_{\o,\kk+\pp}\ra+
\n_-(\pp)
D_{-\o'}(\pp)\la\hat \r_{-\o',\pp}\hp^-_{\o,\kk}\hp^+_{\o,\kk+\pp}\ra\;}\Eq(a4)$$
can be also written as
$$-i\pp_\m
\la j^\m_\pp\psi_\kk\bar\psi_{\kk+\pp}\ra
=[\la\psi_{\kk}\bar\psi_{\kk}\ra
-\la\psi_{\kk+\pp}\bar\psi_{\kk+\pp}\ra]+(\n_+(\pp)+\n_-(\pp))(-i\pp_\m
\la j^\m_\pp\psi_\kk\bar\psi_{\kk+\pp}\ra)
\Eq(a22)$$
$$-i\pp_\m\la j^{5,\m}_\pp\psi_\kk\bar\psi_{\kk+\pp}\ra
=\g^5[\la\psi_{\kk}\bar\psi_{\kk}\ra
-\la\psi_{\kk+\pp}\bar\psi_{\kk+\pp}\ra]+(\n_+(\pp)-\n_-(\pp))
(-i\pp_\m\la j^{5,\m}_\pp\psi_\kk\bar\psi_{\kk+\pp}\ra)
\Eq(a33)$$
\*
We can write $\la\hat \r_{\o',\pp}\hp^-_{\o,\kk}\hp^+_{\o,\kk+\pp}\ra
\equiv\hat G^{2,1}_{\o';\o}(\pp;\kk)$ and  $\la\psi_{\o,\kk}\psi^+_{\o,\kk}\ra\equiv
\hat G^{2}_{\o}(\kk)$ as
a (non convergent) power series in $\l$
$$\hat G^{2,1}_{\o';\o}(\pp;\kk)=\sum_{n=0}^\io \hat G^{2,1(n)}_{\o';\o}(\pp;\kk)\l^n
\quad\quad \hat G^{2}_{\o}(\kk)=\sum_{n=0}^\io \hat G^{2(n)}_{\o}(\kk)\l^n\Eq(a32)$$
The perturbative contributions to 
$\hat G^{2,1}_{\o';\o}(\pp;\kk)$, $\hat G^{2}_{\o}(\kk)$ can be obtained
by a standard Feynman graph expansion with propagator $g_\o^{(\le N)}(\kk)$.
A crucial role will be played by the following identity
$$g_\o^{(\le N)}(\kk)g_\o^{(\le N)}(\kk+\pp)=
{g_\o^{(\le N)}(\kk)-g_\o^{(\le N)}(\kk+\pp)\over D_\o(\pp)}-
g_\o^{(\le N)}(\kk)g_\o^{(\le N)}(\kk+\pp) 
{C_{N}(\kk,\kk+\pp)\over D_\o(\pp)}\Eq(a31)$$
where $C_{N}(\kk,\kk+\pp)$ is given by \equ(ccc).
Of course if $|\kk|\le \g^{N-2}$ the second addend in \equ(a31) is vanishing.
We get
$$\hat G^{2(0)}_{\o}(\kk)=g_\o^{(\le N)}(\kk)
\quad\quad\quad \hat G^{2,1(0)}_{\o;\o}(\pp;\kk)=g_\o^{(\le N)}(\kk)g_\o^{[h,N]}(\kk+\pp)
\Eq(bww)$$
If $\kk,\pp$ are "far" from the cutoffs, that is  
$|\kk|,|\kk+\pp|\le \g^{N-2}$ we get from \equ(a31)
$$\hat G^{2,1(0)}_{\o';\o}(\pp;\kk)=
g_\o^{(\le N)}(\kk)g_\o^{(\le N)}(\kk+\pp)
={g_\o^{(\le N)}(\kk)-g_\o^{(\le N)}(\kk+\pp)\over D_\o(\pp)}=
{G_\o^{2(0)}(\kk)-G_\o^{2(0)}(\kk+\pp)\over D_\o(\pp)}
$$
and we see that \equ(a4) holds with $\n_+^{(0)}=\n_-^{(0)}=0$.
At first order in $\l$
$$\hat G^{2(1)}_{\o}(\kk)=g_\o^{(\le N)}(\kk)g_\o^{(\le N)}(\kk)\int d\kk' g_\o^{(\le N)}(\kk')=0$$
by parity; that is the tadpole contribution is vanishing. Moreover
$\hat G^{2,1(0)}_{\o;\o}(\pp;\kk)=0$
as there is no graph contributing to it.

At the second order in $\l$ we find, if 
$B^{(\le N)}_{-\o}(\kk_1,\kk_2)=g_{-\o}^{(\le N)}(\kk_1)g_{-\o}^{(\le N)}(\kk_2)
v^2(\kk_1-\kk_2)$
$$\hat G^{2(2)}_{\o}(\kk)=g_\o^{(\le N)}(\kk)[\int d\kk_1 \int d\kk_2 
B_{-\o}^{(\le N)}(\kk_1,\kk_2)g_{\o}^{(\le N)}(\kk-\kk_2+\kk_1)]g_\o^{(\le N)}(\kk)\Eq(112)
$$

\*
\insertplot{300pt}{100pt}%
{.
}%
{verticiT11b}{\eqg(1v)}

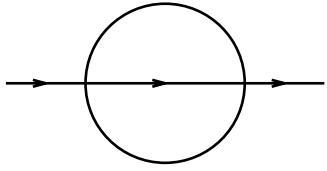
\centerline{Fig 11: Feynmann graph of $\hat G^{2(2)}_{\o}(\kk)$}
\* \vbox{{\ }

}
\*
On the other hand $\hat G^{2,1(2)}_{\o,\o}$ is given by three graphs
$$\hat G^{2,1(2)}_{\o,\o}(\kk,\pp)=
\hat G^{2,1(2)}_{a,\o,\o}(\kk,\pp)+\hat G^{2,1(2)}_{b,\o,\o}(\kk,\pp)
+\hat G^{2,1(2)}_{c,\o,\o}(\kk,\pp)\Eq(aaas)$$
where $\hat G^{2,1(2)}_{a,\o}(\kk,\pp)$ is given by
%
%
\*
\insertplot{300pt}{100pt}%
{.
}%
{verticiT11a}{\eqg(1v)}

\centerline{Fig 12: Feynmann graph of $\hat G^{2,1(2)}_{a,\o}$}
\* \vbox{{\ }
}
\*
while $\hat G^{2,1(2)}_{b,\o}(\kk,\pp)$ is given by
%
\*
\insertplot{300pt}{100pt}%
{.
}%
{verticiT11c}{\eqg(1v)}

\centerline{Fig 13: Feynmann graph of $\hat G^{2,1(2)}_{b,\o}$}
\* \vbox{{\ }
}
\*
and $\hat G^{2,1(2)}_{c,\o}(\kk,\pp)$ is given by
%
\*
\insertplot{300pt}{100pt}%
{.
}%
{verticiT11d}{\eqg(1v)}

\centerline{Fig 14: Feynmann graph of $\hat G^{2,1(2)}_{c,\o}$}
\* \vbox{{\ }
}
\*
We get, using \equ(a31) and if $|\kk|,|\kk+\pp|\le 2^{N-2}$
$$D_\o(\pp)\hat G^{2,1(2)}_{b,\o,\o}(\kk,\pp)-
\hat G^{2(2)}_{\o}(\kk)+
\hat G^{2(2)}_{\o}(\kk+\pp)=$$
$$g_{\o}^{(\le N)}(\kk)\{\int d\kk_1 d\kk_2 
B_{-\o}^{(\le N)}(\kk_1,\kk_2)
[g_{\o}^{(\le N)}(\kk+\pp-\kk_2+\kk_1)-
g_{\o}^{(\le N)}(\kk-\kk_2+\kk_1)
]\}
g_{\o}^{(\le N)}(\kk+\pp)\Eq(jhjhn)$$
and using again \equ(a31) the r.h.s. of \equ(jhjhn) can be written
as
$$-D_\o(\pp)\hat G^{2,1(2)}_{a,\o,\o}(\kk,\pp)+
g_\o^{(\le N)}(\kk)g_\o^{(\le N)}(\kk+\pp)
[\int d\kk_1 d\kk_2 
B_{-\o}^{(\le N)} g_{\o}^{(\le N)}(\kk+\pp-\kk_2+\kk_1)
g_\o^{(\le N)}(\kk-\kk_2+\kk_1)C_{N,\o}
]$$
Finally, by using again \equ(a31) and the fact that, by parity
$\int d\kk g_{\o}^{(\le N)}(\kk)=0$, we get
$$\hat G^{2,1(2)}_{c,\o}(\kk,\pp)=[\int d\kk_1 
g_{\o}^{(\le N)}(\kk_1)
g_\o^{(\le N)}(\kk_1+\pp)
C_{h,N}(\kk_1,\kk_1+\pp)]
[\int d\kk_2 g_{-\o}^{(\le N)}(\kk_2)
g_{-\o}^{(\le N)}(\kk_2+\pp)]g_\o^{(\le N)}(\kk)g_\o^{(\le N)}(\kk+\pp)]$$
and,using that
$$\hat G^{2,1(1)}_{\o,-\o}(\kk,\pp)=
\int d\kk_2 g_{-\o}^{(\le N)}(\kk_2)
g_{-\o}^{(\le N)}(\kk_2+\pp)$$
we find at the end,putting togheter all terms
$$D_\o(\pp)\hat G^{2,1(2)}_{\o,\o}(\kk,\pp)=
\hat G^{2(2)}_{\o}(\kk)-\hat G^{2(2)}_{\o}(\kk+\pp)+
\tilde\n_-^{(1)}(\kk,\pp)D_{-\o}(\pp)\hat G^{2,1(1)}_{-\o,\o}(\kk,\pp)+
\tilde\n_+^{(2)}(\kk,\pp)D_\o(\pp)\hat G^{2,1(0)}_{\o,\o}(\kk,\pp)\Eq(ddff)$$
where
$$\tilde\n_-^{(1)}(\kk,\pp)=\int d\kk_1 
g_{\o}^{(\le N)}(\kk_1)
g_\o^{(\le N)}(\kk_1+\pp)
{C_{\o,N}(\kk_1,\kk_1+\pp)\over D_{-\o}(\pp)}\Eq(fon1)$$
$$\tilde\n_+^{(2)}(\kk,\pp)=\int d\kk_1 \int d\kk_2 v^2(\kk_1-\kk_2)
g_{-\o}^{(\le N)}(\kk_1)g_{-\o}^{(\le N)}(\kk_2)g_{\o}^{(\le N)}(\kk+\pp-\kk_2+\kk_1)
g_\o^{(\le N)}(\kk-\kk_2+\kk_1){C_{\o,N}\over D_{\o}(\pp)}$$
Note that $\tilde\n_-^{(1)}(\pp)$ coincides with \equ(fhnk).
On the other hand the value of $\tilde\n_+^{(2)}(\pp)$
depends crucially on  $v(\pp)$. 

If $v(\pp)$ decays for large $\pp$,
using \equ(mjmj)
we can write 
$$\tilde\n_+^{(2)}(\kk,\pp)=\int d\kk'\sum_{h=-\io}^N 
v^2(\kk-\kk')[\D^{h,N}
(\kk',\kk'+\pp)+\D^{N,h}
(\kk',\kk'+\pp)]
 A(\kk-\kk'')$$
where
$$A(\pp)
=\int d\kk''
 g_{-\o}^{(\le N)}(\kk'')g_{-\o}^{(\le N)}(\pp-\kk'')$$
As $\sup_\pp |A(\pp)|\le C$ (it is the same as $\n_-$ using 
\equ(a31) and noting that $\int d\kk g$ is vanishing)
we get, remembering
that $\kk,\pp$ are fixed and $|v(\pp)|\le C (\pp^2+1)^{-1}$ and if $C_{\kk,\pp}$ 
is a $\kk,\pp$-dependent constant
$$|\tilde\n_+^{(2)}(\pp)|\le C_{\kk,\pp} \sum_{h=-\io}^N \l {1\over \g^{2 h}+1}
\g^{h}\g^{-N}\le \l \bar C_{\kk,\pp} \g^{-N}$$
which is vanishing for $N\to\io$.

On the other hand if $v(\pp)=1$ we get,for $\kk,\pp=0$
$$\n_+^{(2)}(0)=\int {d\kk\over (2\pi)^2}
\left[{u_0(|\kk|)\chi_0(|\kk|)\over
|\kk|^4}- {\chi'_0(|\kk|)\over
2|\kk|^3}\right]
A(\kk)D^2_{-\o}(\kk)$$
which is {\it nonvanishing};
on the other hand the rest is vanishing as $N\to\io$ as $\pp,\kk$
are fixed,by dimensional reasons.

\*
\vskip1cm
{\bf Acknowledgments.} I am deeply indebted with K. Gawedzki
for very illuminating discussions.
\*
{\bf\centerline{\titolo References}}
\*
\halign{\hbox {#\hss} & \vtop{\advance\hsize by -0.8
truecm \0#}\cr
%
%
[A]& {Adler S. L.: {\journal Phys. Rev.}, {\bf 177},
{\pagine 2426--2438} 1969.}\cr\cr
[A1]& {Adler S. L.: hep-th/0405040
}\cr\cr
[AB] & {Adler S. L., Bardeen W.A.: {\journal Phys. Rev.}
{\bf 182}, {\pagine 1517-1536} 1969.}\cr\cr
%
%
[AI] & {Anselm A. , Iogansen A.:
{\journal Sov. Phys. JETP} {\bf 96}, 
{\pagine 670-62} 1989.}\cr\cr
[B] &  {R. Baxter, Exactly solved models in statistical mechanics,
Academic Press 1982}\cr\cr
[BFM] & {Benfatto G., Falco P, Mastropietro V.: hep-th0606177 
}\cr\cr
[BM]& {Benfatto G., Mastropietro V.:
{\journal Rev. Math. Phys.}, {\bf 13},
{\pagine 1323--1435}, 2001;
{\journal Comm. Math. Phys.}, {\bf 231},
{\pagine 97--134}, 2002.;{\journal Comm. Math. Phys.}, {\bf 258},
{\pagine 609--655}, 2005. }\cr\cr
%
%
%
%
%
%
[DMT]& {DeRaad L.L., Milton K.A., Tsai W.:
{\journal Phys. Rev.D}, {\bf 6},
{\pagine 1766--1780}, 1972. }\cr\cr
%
%
[F]& {Fujikawa K.:
{\journal Phys. Rev. Lett.}, {\bf 42}
{\pagine 1195--1198}, 1979. }\cr\cr
[GK]& {Gawedzki K, Kupiainen, A.:
{\journal Comm. Math.Phys.}, {\bf 102},
{\pagine 1--30}, 1985. }\cr\cr
%
%
[G]&{G.Gallavotti. {\it Rev. Mod.Phys.} 55, 471, (1985) }\cr\cr
[GL]& {Gomes M., Lowenstein J.H.:
{\journal Nucl. Phys. B}, {\bf 45},
{\pagine 252--266}, 1972. }\cr\cr
[GM]& {Gentile G, Mastropietro V.:
{\journal Phys Rep} 352, 4-6,
{\pagine 273-343} (2001) }\cr\cr
[GM1]& {Giuliani A, Mastropietro V.:
{\journal Phys Rev Lett} 93,19 (2004)}\cr\cr
[GR]& {Georgi H, Rawls J {\journal Phys. Rev. D} 3,4,874--879}\cr\cr 
%
%
[J]& {Johnson K.:
{\journal Nuovo Cimento}, {\bf 20},
{\pagine 773--790}, 1961. }\cr\cr
[JJ]& {Jackiw K. Johnson J:
{\journal Phys. Rev. }, {\bf 182},
{\pagine 1459--11469}, 1969. }\cr\cr
%
%
[K]& {Klaiber B.: Lectures in theoretical physics. 
ed: A.O. Barut and W.E. Brittin. {\it
Gordon and Breach.},  1968.}\cr\cr
[Le]& {A. Lesniewski: {\it Commun. Math. Phys.} {\bf 108}, 437--467
1987. }\cr\cr
[LS]& {Lowenstein J.H, Schroer B: 
{\it Phys. Rev. D} {\bf 7}, 1929--1933
1973. }\cr\cr
%
%
%
%
[P] &{ Polchinski,J. {\it Nucl. Phys. B} 231 , 269 (1984)
}\cr\cr
[Se]& {Seiler E.:  {\it Phys Rev D} 22, 2412(1980). }\cr\cr
%
%
%
%
%
%
[Z]& {Zee A. 
{\it Phys. Rev. Lett.} 1198 (1972). 
}\cr\cr
%
}

\bye